\documentclass[aps,pre,twocolumn,groupedaddress]{revtex4-2}

\usepackage{amsmath,amssymb,bm}
\usepackage{graphicx,xcolor}
\usepackage{tikz}

\newcommand{\E}{\mathbb{E}}                 %
\newcommand{\Var}{\operatorname{Var}}       %
\newcommand{\diag}{\operatorname{diag}}

\newcommand{\vx}{x_A}                       %
\newcommand{\vy}{x_B}                       %
\newcommand{\vX}{X_A}                       %
\newcommand{\vY}{X_B}
\newcommand{\PA}{P_A}                       %
\newcommand{\PB}{P_B}                       %

\newcommand{\rhoHGR}{\rho_{\mathrm{HGR}}}

\newcommand{\T}{\mathsf{T}}                 %
\newcommand{\Rr}{\mathbb{R}}

\usetikzlibrary{positioning,decorations.pathreplacing}
\tikzset{
  core/.style={circle,draw,thick,inner sep=0pt,minimum size=3.4mm},
  beta/.style={core,fill=orange!60},
  hole/.style={core,dashed,draw=orange!70!black},
  ones/.style={core,font=\scriptsize},
  bond/.style={thick},
  leg/.style={thick},
  brace/.style={decorate,decoration={brace,amplitude=4pt}},
  mbrace/.style={decorate,decoration={brace,mirror,amplitude=4pt}},
}
\newcommand{\Pchain}[2]{%
  \foreach \i in {0,...,4}{\node[core] (#1\i) at (\i,#2) {};}
  \foreach \i in {0,...,3}{\draw[bond] (#1\i)--(#1\the\numexpr\i+1\relax);}
}
\tikzset{pcore/.style={core,fill=blue!20}}
\renewcommand{\Pchain}[2]{%
  \foreach \i in {0,...,4}{\node[pcore] (#1\i) at (\i,#2) {};}
  \foreach \i in {0,...,3}{\draw[bond] (#1\i)--(#1\the\numexpr\i+1\relax);}
}

\begin{document}

\title{Maximal correlation and optimal transformations of probability distributions
in tensor-train format: a variational algorithm}

\author{Kayo Kinjo}
\email{kinjok@mail.saitama-u.ac.jp}
\author{Jun Ohkubo}
\affiliation{Graduate School of Science and Engineering, Saitama University, Sakura, Saitama, 338–8570 Japan}

\date{\today}

\begin{abstract}
The correlation between two subsystems $A$ and $B$ of a many-body system
is usually measured through a pair of observables chosen in advance.  The
Hirschfeld--Gebelein--R\'enyi maximal correlation removes the choice: it is
the largest correlation coefficient between any function of the
configuration of $A$ and any function of the configuration of $B$, and the
two functions that reach it, the optimal transformations, are the
observables that carry the correlation.  Both follow from a singular value
decomposition (SVD) of the joint distribution normalized by its marginals,
a matrix whose dimension is the number of configurations.  We show that
this SVD can be computed variationally for a classical distribution of
discrete variables, such as the Boltzmann distribution of a spin model,
given in the tensor-train (matrix-product) format, without ever forming
the matrix.  The optimal transformations are sought as tensor trains; each
step updates a single core in closed form, and the inverse of a small
matrix contracted from the distribution and the remaining cores performs
the division by the marginals core by core.  Every iterate is a feasible
pair, so the objective is a lower bound on the maximal correlation that
increases monotonically, and the cost grows linearly with the number of
variables.  The algorithm reproduces the exact SVD on small systems,
recovers the known nonlinear optimal transformations of synthetic
distributions of up to ten angular variables ($10^{15}$ configurations),
and, for the bipartite antiferromagnetic model with $2^{32}$
configurations, returns an optimal transformation that is a function of
the group magnetization, linear above the transition and sign-like below
it: the staggered order parameter is read off from the distribution
instead of being postulated.
\end{abstract}

\maketitle

\section{Introduction}
\label{sec:intro}

The correlation between two subsystems $A$ and $B$ of a classical
many-body system is usually measured through observables that are chosen
in advance, one on each subsystem: a function $O_A$ of the configuration
of $A$ and a function $O_B$ of the configuration of $B$.  A correlation
function $\langle O_AO_B\rangle$ measures only the correlation carried by
the pair $(O_A,O_B)$ chosen beforehand; if the pair is poorly chosen, the
correlation looks weak even when $A$ and $B$ are strongly dependent.
The mutual information avoids the choice altogether: it
measures the total dependence between $A$ and $B$ whatever carries it, but
for the same reason it returns no observable. This paper starts from the
question that lies between the two. Given the joint distribution of $A$ and
$B$, can the pair of observables that carries the correlation be extracted
from the distribution itself, instead of being postulated?

For distributions of a few variables, statistics has had an answer to this
question since 1935. The Hirschfeld--Gebelein--R\'enyi (HGR) maximal
correlation \cite{Hirschfeld1935,Gebelein1941,Renyi1959} is the largest
correlation coefficient that can be reached between any function of the
configuration of $A$ and any function of the configuration of $B$. These
two functions are called the optimal transformations
in Ref.~\cite{Breiman1985}, are precisely the observables that carry the strongest
correlation. 
Both are obtained from a singular value decomposition (SVD) of the distribution itself
\cite{Hirschfeld1935,Lancaster1958,Greenacre1984,Huang2019}. The matrix in
question is the joint distribution $P(\vx,\vy)$, with row index $\vx$ and
column index $\vy$, normalized by the square root of the product of its
marginals $\PA(\vx)$ and $\PB(\vy)$, $K=P/\sqrt{\PA\PB}$.
The largest singular value of $K$ is
one and its singular vectors correspond to the constant functions on $A$ and $B$, respectively. 
The second largest one is the HGR maximal
correlation. Its singular vectors are the optimal transformations, and the
singular values after it are the largest correlations that remain once the
earlier pairs of functions have been excluded.
For a Gibbs distribution, one expects the optimal transformation to be a function
of the order parameter, so that the order parameter comes out of the SVD
instead of going in. The problem is thus one of linear algebra, a
singular value problem for $K$.  In the language of statistics it is a
canonical correlation analysis \cite{Hotelling1936}, but one taken over
the indicator functions of the configurations of $A$ and of $B$ rather
than over a chosen set of variables; since the indicators span all
functions of the configuration, this is what makes the transformations
nonlinear and the object of the analysis the distribution itself.  Its
single difficulty is that the dimension of $K$ is the number of
configurations, exponential in the system size.

Vectors and matrices of exactly this size are the daily business of
tensor-network methods. A classical distribution over $N$ discrete
variables, such as a spin model, can be
stored as a tensor train (TT) \cite{Oseledets2011}, the format known in
physics as a matrix product state (MPS); the Boltzmann weight of a chain, a
product of transfer matrices, is the simplest example. Physicists also know
an SVD of such a distribution, but it is a different one. The SVD of an MPS
at a bond is the Schmidt decomposition \cite{Schollwoeck2011}, and it
has been applied to classical distributions as well, through $\sqrt{P}$ or
directly \cite{RokhsarKivelson1988,LiHaldane2008,TemmeVerstraete2010}. This
bond SVD decomposes the \emph{state} into products of states across the cut,
and what it measures is how large a bond dimension the state needs there.
The SVD of the previous paragraph decomposes the distribution into products
of \emph{observables}. For a quantum state the two coincide, because a
Schmidt vector is at once a state and, as a projector, an observable;
for a classical system they do not:
the state $P$ is normalized in $L_1$, an
observable $f(\vx)$ is a function of the configuration, and
the two meet only in the expectation value, $\E[f]:=\sum_{\vx}\PA(\vx)f(\vx)$.
Decomposing $P$ into products of distributions is what tensor-network
methods have done so far~\cite{RokhsarKivelson1988,LiHaldane2008,TemmeVerstraete2010};
decomposing it into products of observables has
not been carried out in the TT format.

At this point, it is natural to ask why the SVD of $K$ should not simply be
taken on the TT of $P$: a few extreme singular values and vectors of a
matrix that is itself given in the TT format can be computed by alternating
sweeps \cite{LeeCichocki2015}.  The obstacle is that $K$ is not available in
that format.  The division by the marginals is not a TT operation: $\PA(\vx)$ depends on all
the variables of $A$, and neither its inverse square root nor $K$ has, in
general, a TT representation of small bond dimension, even when $P$ does.
Nor do the methods of statistics help directly, 
since they estimate the optimal transformations
from samples \cite{Breiman1985,Buja1990,Huang2019} and, for a function of
many variables, need either a number of samples that grows with the number
of configurations or a parametric ansatz chosen in advance, such as the
neural networks with which the leading singular functions of an operator
are fitted from samples in Ref.~\cite{Ryu2024}. For a many-body
distribution, the optimal transformations have therefore never been computed to our knowledge.

In this study, we show that the SVD of $K$ can be computed
variationally on the TT of $P$, without ever forming $K$. Each
step of the algorithm updates a single TT core of $f$ or $g$ in closed form,
and the inverse of a small matrix, contracted from the distribution and the
remaining cores, performs the division by the marginal core by core.
The algorithm stands to the exact SVD of $K$ as the
density-matrix renormalization group \cite{White1992,Schollwoeck2011} stands
to exact diagonalization: the objective is a lower bound on the HGR maximal
correlation that increases at every sweep, and the cost grows linearly with
$N$. We found that the algorithm reproduces the exact SVD on small systems,
recovers the known nonlinear optimal transformations of synthetic
distributions, and, for a mean-field antiferromagnet, returns an optimal
transformation that is a function of the staggered magnetization. The result
shows that the pair of observables carrying the correlation between two
subsystems of a many-body system can be read off from its distribution,
without being chosen beforehand.

The paper is organized as follows. Section~\ref{sec:hgr} recalls the HGR
problem, shows that it is equivalent to the SVD of $K$, and derives the
self-consistency relations that the optimal transformations satisfy
under the alternation between $f$ and $g$, which is the alternating
conditional expectations (ACE) of statistics \cite{Breiman1985}.
Section~\ref{sec:tt} derives the TT algorithm, the closed-form local
solution, and its convergence properties.  Sections~\ref{sec:synthetic}
and~\ref{sec:afm} apply the algorithm to synthetic distributions of
angles with a known nonlinear optimal transformation and to the
bipartite mean-field antiferromagnet; in both cases the exact reference
is obtained by a reduction to sufficient statistics
(Appendix~\ref{app:suff}), and on small systems the algorithm is also
compared with the exact SVD of $K$ (Appendix~\ref{app:svdcheck}).

\section{HGR maximal correlation and the singular value decomposition}
\label{sec:hgr}

\subsection{Definition}
\label{sec:def}

Let $\vX$ and $\vY$ be discrete random variables taking values $\vx\in
\mathcal X_A$ and $\vy\in\mathcal X_B$ with joint distribution $P(\vx,\vy)$
and marginals $\PA(\vx)=\sum_{\vy}P(\vx,\vy)$ and $\PB(\vy)=\sum_{\vx}
P(\vx,\vy)$.  Throughout we assume that $\mathcal X_A$ and $\mathcal X_B$
contain only configurations of nonzero marginal probability, $\PA(\vx)>0$
and $\PB(\vy)>0$; individual joint probabilities $P(\vx,\vy)$ may vanish.

The HGR maximal correlation $\rhoHGR$ is defined by
\cite{Hirschfeld1935,Gebelein1941,Renyi1959}
\begin{align}
  \rhoHGR &=\max_{f,g}\;\E\!\left[f(\vX)\,g(\vY)\right],
  \label{eq:hgr}\\
  \text{s.t.}\quad
  &\E[f(\vX)]=0,\qquad \E[g(\vY)]=0,
  \label{eq:const1}\\
  &\Var[f(\vX)]=1,\qquad \Var[g(\vY)]=1,
  \label{eq:const2}
\end{align}
where $f:\mathcal X_A\to\Rr$ and $g:\mathcal X_B\to\Rr$ are arbitrary
(in general nonlinear) functions.  The maximizers $(f^{*},g^{*})$ are
called the optimal transformations \cite{Breiman1985}.  By the
Cauchy--Schwarz inequality $0\le\rhoHGR\le 1$; the lower bound holds because
$(f,g)$ and $(f,-g)$ are both feasible.

\subsection{Equivalence to a singular value decomposition}
\label{sec:svd}

In this subsection we show that finding the optimal transformations is
equivalent to computing the SVD of a matrix associated with the joint
distribution.  First, the constraints \eqref{eq:const1}--\eqref{eq:const2}
are expressed through inner products weighted by the marginals.  For
functions $f,f'$ on $\mathcal X_A$, we denote the inner product weighted by
$\PA$ as
\begin{equation}
  \langle f,f'\rangle_{\PA}:=\sum_{\vx}\PA(\vx)f(\vx)f'(\vx)=\E[f(\vX)f'(\vX)],
  \label{eq:winner}
\end{equation}
and likewise $\langle\cdot,\cdot\rangle_{\PB}$ on $\mathcal X_B$. The constraints become
\begin{equation}
  \E[f(\vX)]=\langle f,1\rangle_{\PA}=0,\quad
  \Var[f(\vX)]=\langle f,f\rangle_{\PA}=1,
  \label{eq:const_inner}
\end{equation}
so that the feasible $f$ are the unit vectors orthogonal to the constant
function in the geometry of Eq.~\eqref{eq:winner}, and similarly for $g$.

The objective, on the other hand, is a bilinear form in $f$ and $g$.
Regarding $f$ and $g$ as column vectors indexed by the configurations of
$A$ and $B$, and $P$ as the matrix whose rows are labeled by $\vx$ and
whose columns by $\vy$,
\begin{equation}
  \E[f(\vX)g(\vY)]=\sum_{\vx,\vy}f(\vx)P(\vx,\vy)g(\vy)=f^{\T}Pg .
  \label{eq:bilinear_P}
\end{equation}
The problem is therefore the maximization of a bilinear form over pairs of
unit vectors that are orthogonal to the constant function, which is the
variational characterization of a singular value.  The only difference from
the usual setting is that the relevant inner products are the weighted ones
of Eq.~\eqref{eq:winner} rather than the Euclidean ones in which the SVD is
defined.

The two are related by a change of coordinates.  Let $D_A=\diag(\PA)$,
$D_B=\diag(\PB)$ and rescale
\begin{equation}
  \begin{split}
  &u=D_A^{1/2}f,\qquad v=D_B^{1/2}g,\qquad\text{i.e.}\\
  &u(\vx)=\sqrt{\PA(\vx)}\,f(\vx),\quad v(\vy)=\sqrt{\PB(\vy)}\,g(\vy).
  \end{split}
  \label{eq:rescale}
\end{equation}
Then $\langle f,f'\rangle_{\PA}=u^{\T}u'$, i.e., the weighted inner product
becomes the Euclidean one, and the objective~\eqref{eq:bilinear_P} becomes
\begin{equation}
  \begin{split}
  f^{\T}Pg&=u^{\T}Kv,\qquad K:=D_A^{-1/2}\,P\,D_B^{-1/2},\\
  K(\vx,\vy)&=\frac{P(\vx,\vy)}{\sqrt{\PA(\vx)\,\PB(\vy)}} .
  \end{split}
  \label{eq:K}
\end{equation}
Thus $K$ is nothing but the distribution $P$ written in orthonormal
coordinates for the weighted inner products; we call it the
\emph{normalized distribution}.  In statistics its SVD underlies
correspondence analysis \cite{Hirschfeld1935,Greenacre1984}; in information
theory $K$ is known as the divergence transfer matrix \cite{Huang2019}.

Let
\begin{equation}
  K=U\Sigma V^{\T}=\sum_{i\ge 1}\sigma_i\,u_i\,v_i^{\T},
  \qquad
  \sigma_1\ge\sigma_2\ge\cdots\ge 0 ,
  \label{eq:svd}
\end{equation}
be the SVD of $K$, with orthonormal singular vectors $\{u_i\}$ and
$\{v_i\}$.  
Since the zero-mean constraint in Eq.~\eqref{eq:const_inner} is the
orthogonality to the constant function, we first look at where the constant
functions go under the rescaling in Eq.~\eqref{eq:rescale}.  They become
\begin{equation}
  u_1:=D_A^{1/2}\,1=\sqrt{\PA},\qquad v_1:=D_B^{1/2}\,1=\sqrt{\PB},
  \label{eq:u1}
\end{equation}
which are unit vectors, since $\sum_{\vx}\PA(\vx)=1$.  Moreover, summing
$P$ over $\vy$ gives the marginal $\PA$, so that
\begin{equation}
  (Kv_1)(\vx)=\sum_{\vy}\frac{P(\vx,\vy)}{\sqrt{\PA(\vx)}}
  =\frac{\PA(\vx)}{\sqrt{\PA(\vx)}}=u_1(\vx),
  \label{eq:trivial}
\end{equation}
and likewise $K^{\T}u_1=v_1$ by summing over $\vx$.
Hence $(u_1,v_1)$ is a singular pair of $K$ with singular value one.  This
is the largest singular value: for any unit vectors $u,v$, Eq.~\eqref{eq:K}
and the Cauchy--Schwarz inequality give $u^{\T}Kv=\E[fg]\le
\sqrt{\E[f^2]\E[g^2]}=1$.  Thus $\sigma_1=1$, and we call $(u_1,v_1)$ the
trivial mode\footnote{If $\sigma_2=1$ as well (perfect correlation),
the singular vectors for the singular value one
are not unique, and we fix the basis of that subspace so that
$u_1=\sqrt{\PA}$ and $v_1=\sqrt{\PB}$ belong to it;
Eqs.~\eqref{eq:rho=sigma2}--\eqref{eq:fstar} hold with this choice.}.
The zero-mean constraint $\langle f,1\rangle_{\PA}=0$ is exactly the
condition $u\perp u_1$: it removes the trivial mode.  Therefore
\begin{equation}
  \rhoHGR=\max_{\substack{\|u\|=\|v\|=1\\ u\perp u_1,\;v\perp v_1}}
  u^{\T}Kv=\sigma_2.
  \label{eq:rho=sigma2}
\end{equation}
The proof is
given in Appendix~\ref{app:proof}.  By Eq.~\eqref{eq:rescale} the
optimal transformations are
\begin{equation}
  f^{*}(\vx)=\frac{u_2(\vx)}{\sqrt{\PA(\vx)}},\qquad
  g^{*}(\vy)=\frac{v_2(\vy)}{\sqrt{\PB(\vy)}} .
  \label{eq:fstar}
\end{equation}

More generally, $f_i=u_i/\sqrt{\PA}$ and $g_i=v_i/\sqrt{\PB}$ ($i\ge 1$,
with $f_1=g_1=1$) are orthonormal in the weighted inner products, and the
SVD in Eq.~\eqref{eq:svd} is equivalent to the canonical expansion of Lancaster
\cite{Lancaster1958},
\begin{equation}
  P(\vx,\vy)=\PA(\vx)\PB(\vy)\Bigl[1+\sum_{i\ge 2}\sigma_i f_i(\vx)g_i(\vy)\Bigr],
  \label{eq:lancaster}
\end{equation}
which separates $P$ into its independent part $\PA\PB$ and a sum of
rank-one corrections ordered by strength.  Each $\sigma_i$ is the ordinary
Pearson correlation coefficient of the pair of standardized statistics
$(f_i(\vX),g_i(\vY))$, and different pairs are uncorrelated, that is,
$\E[f_i g_j]=\sigma_i\delta_{ij}$.
We refer to the sequence $1=\sigma_1\ge\sigma_2=\rhoHGR\ge\sigma_3\ge\cdots$
as the \emph{dependence spectrum}; the trivial mode is always counted as
$\sigma_1=1$, so that $\rhoHGR=\sigma_2$ throughout.
The same sequence appears under other names: the squares
$\sigma_i^2$ ($i\ge2$) are the principal inertia components of
Ref.~\cite{Calmon2017}, the word inertia coming from correspondence
analysis \cite{Greenacre1984}, and the expansion in Eq.~\eqref{eq:lancaster}
is the modal decomposition of $P$ in Ref.~\cite{Huang2019}.
Note that $\sigma_i=0$ for all $i\ge 2$ if and only if $P=\PA\PB$, i.e., 
$\rhoHGR=0$ if and only if $\vX$ and $\vY$ are independent.

\subsection{Self-consistency relations for the optimal transformations}
\label{sec:fixedpoint}

The SVD of Sec.~\ref{sec:svd} treats $f$ and $g$ at once.  The optimal
transformations can also be characterized one at a time, by what happens
when one of them is held fixed: the maximization over the other can then
be done in closed form, and the optimal pair $(f^{*},g^{*})$ is reproduced
by these updates.  These relations are the basis of the alternating
algorithm developed in this study.

Consider first the maximization over $g$ alone, for a fixed
$f$.  By the tower property of conditional expectation, the objective
depends on $f$ only through the conditional expectation of $f(\vX)$ given
$\vY$:
\begin{align}
  \E[f(\vX)g(\vY)]&=\E_{\vY}\bigl[g(\vY)\,h_f(\vY)\bigr],
  \label{eq:tower}\\
  h_f(\vy)&:=\E_{\vX}[f(\vX)\,|\,\vY=\vy]
  \nonumber\\
  &\;=\sum_{\vx}\frac{P(\vx,\vy)}{\PB(\vy)}f(\vx).
  \label{eq:hf}
\end{align}
The function $h_f$ is automatically centered, since
$\E[h_f(\vY)]=\E[f(\vX)]=0$.
For $g$ with $\E[g]=0$ and $\Var[g]=1$, the Cauchy--Schwarz inequality gives
\begin{equation}
  \E\bigl[g\,h_f\bigr]\le\sqrt{\Var[g]\,\Var[h_f]}=\sqrt{\Var[h_f(\vY)]},
\end{equation}
with equality if and only if $g$ is proportional to $h_f$; together with
$\E[g^2]=1$ this gives
\begin{equation}
  g(\vy)=\frac{h_f(\vy)}{\sqrt{\Var[h_f(\vY)]}} .
  \label{eq:gopt}
\end{equation}
Thus, for fixed $f$, the optimal $g$ is the \emph{normalized conditional
expectation} of $f$, and the resulting value of the objective is
$\sqrt{\Var[h_f]}$.  Note that $\Var[h_f]\le\Var[f]=1$ by the law of
total variance, $\Var[f]=\E[\Var[f|\vY]]+\Var[\E[f|\vY]]$, with equality
if and only if $f(\vX)$ is a function of $\vY$. Conditioning can only
reduce the variance.

At the optimum, $g^{*}$ is the best $g$ for the fixed
$f^{*}$, so Eq.~\eqref{eq:gopt} holds with $f=f^{*}$, and the value
$\sqrt{\Var[h_{f^{*}}]}$ equals $\rhoHGR$.  Hence
$\E[f^{*}(\vX)|\vY]=h_{f^{*}}(\vY)=\rhoHGR\,g^{*}(\vY)$ and, by symmetry,
\begin{equation}
  \begin{split}
  \E_{\vX}\bigl[f^{*}(\vX)\,\big|\,\vY\bigr]&=\rhoHGR\,g^{*}(\vY),\\
  \E_{\vY}\bigl[g^{*}(\vY)\,\big|\,\vX\bigr]&=\rhoHGR\,f^{*}(\vX).
  \end{split}
  \label{eq:fixedpoint}
\end{equation}
Equation~\eqref{eq:fixedpoint} states that the optimal pair is reproduced
by the $g$-update in Eq.~\eqref{eq:gopt}: the conditional expectation
$\E[f^{*}|\vY]$ of the unit-variance function $f^{*}$ has variance
$\rhoHGR^{2}\le 1$, and after division by $\rhoHGR$ it is exactly $g^{*}$;
in the same way $g^{*}$ is mapped back to $f^{*}$.  In other words,
$(f^{*},g^{*})$ is a fixed point of the alternating conditional expectations
(ACE) algorithm of Breiman and Friedman \cite{Breiman1985,Buja1990}, which
repeats the $g$-update and the $f$-update in turn.
In the matrix language of Sec.~\ref{sec:svd}, Eq.~\eqref{eq:fixedpoint}
reads $K^{\T}u_2=\sigma_2v_2$ and $Kv_2=\sigma_2u_2$, respectively
[cf.\ Eqs.~\eqref{eq:rescale} and~\eqref{eq:fstar}]: the two
conditional-expectation operators are represented by $K^{\T}$ and $K$.
Composing the two relations,
\begin{equation}
  \E\Bigl[\E\bigl[g^{*}(\vY)\,\big|\,\vX\bigr]\,\Big|\,\vY\Bigr]
  =\rhoHGR^2\,g^{*}(\vY),
  \label{eq:composed}
\end{equation}
which identifies $g^{*}$ as an eigenfunction, with eigenvalue $\rhoHGR^2$,
of the composed conditional-expectation operator
$g\mapsto\E[\E[g|\vX]\,|\,\vY]$, represented by $K^{\T}K$ in the coordinates
of Eq.~\eqref{eq:rescale}; likewise $f^{*}$ is an eigenfunction of
$f\mapsto\E[\E[f|\vY]\,|\,\vX]$, represented by $KK^{\T}$.
The convergence of the alternation to $(f^{*},g^{*})$, at a rate set by
the ratio $\sigma_3/\sigma_2$, is shown in Appendix~\ref{app:power}.

ACE evaluates the conditional expectations in one of two ways.  When
the distribution is given as a table, the sum in Eq.~\eqref{eq:hf} is
carried out over the table; this is the form in which the alternation is
used in Ref.~\cite{Huang2019}, and in the network maximal correlation of
Ref.~\cite{Feizi2017}, which runs the same alternation between several
discrete variables at once.  When only a data sample
$\{(\vx^{(s)},\vy^{(s)})\}_{s=1}^{M}$ is available, as in the original
ACE, the conditional expectations are
estimated by
nonparametric regression, i.e.,  by averaging $f(\vx^{(s)})$ over the sample
points whose $\vy^{(s)}$ lie close to $\vy$ (the data smooths of
Ref.~\cite{Breiman1985}), and the transformation of a multivariate
predictor is restricted to the additive form $\sum_i\phi_i(x_i)$.  For a
many-body distribution neither route is available; the exact sum in
Eq.~\eqref{eq:hf} runs over all configurations of $A$, whose number grows
exponentially with the number of variables, and a sample-based estimate
of a function on that space would require a number of samples growing in
the same way.  In the next section, the alternation between $f$ and $g$
is kept, but each update is evaluated exactly on the TT representation of
$P$, with $f$ and $g$ themselves in the TT format.

\section{Variational algorithm in the tensor-train format}
\label{sec:tt}

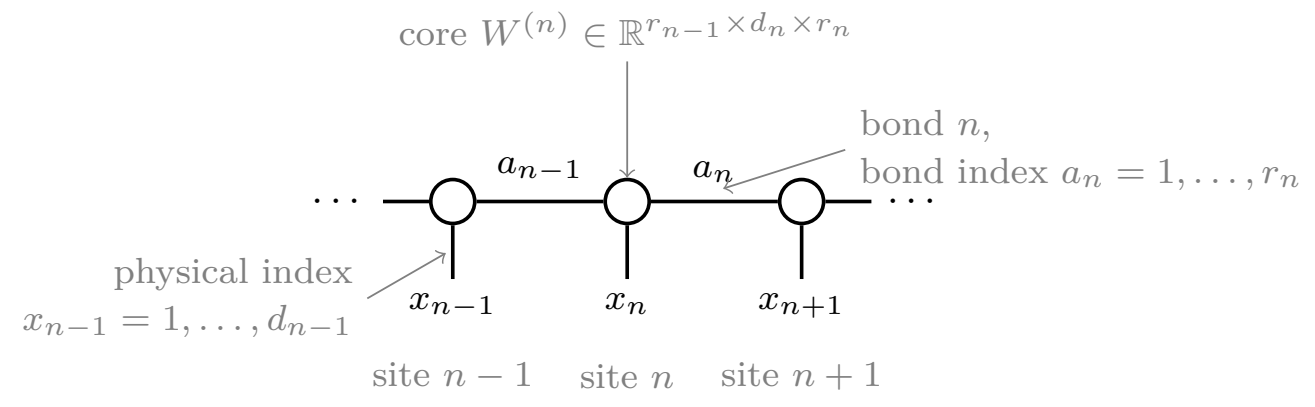
\begin{figure}
\centering
\resizebox{\columnwidth}{!}{%
\begin{tikzpicture}[x=1.35cm,y=1cm]
  \node at (0.35,0) {$\cdots$};
  \node at (3.65,0) {$\cdots$};
  \foreach \i in {1,2,3}{\node[core] (W\i) at (\i,0) {};}
  \draw[bond] (0.6,0)--(W1);
  \draw[bond] (W1)--(W2) node[midway,above=1pt]{\footnotesize $a_{n-1}$};
  \draw[bond] (W2)--(W3) node[midway,above=1pt]{\footnotesize $a_{n}$};
  \draw[bond] (W3)--(3.4,0);
  \foreach \i in {1,2,3}{\draw[leg] (W\i)--++(0,-0.6);}
  \node[below] at (1,-0.6) {\footnotesize $x_{n-1}$};
  \node[below] at (2,-0.6) {\footnotesize $x_{n}$};
  \node[below] at (3,-0.6) {\footnotesize $x_{n+1}$};
  \node[text=gray] at (1,-1.35) {\footnotesize site $n-1$};
  \node[text=gray] at (2,-1.35) {\footnotesize site $n$};
  \node[text=gray] at (3,-1.35) {\footnotesize site $n+1$};
  \draw[<-,gray] (W2.north) -- ++(0,0.9)
    node[above]{\footnotesize core $W^{(n)}\in\Rr^{r_{n-1}\times d_n\times r_n}$};
  \draw[<-,gray] (2.55,0.1) -- ++(0.7,0.3)
    node[right,align=left]{\footnotesize bond $n$,\\ \footnotesize bond index $a_n=1,\dots,r_n$};
  \draw[<-,gray] (0.96,-0.4) -- ++(-0.45,-0.35)
    node[left,align=right]{\footnotesize physical index\\ \footnotesize $x_{n-1}=1,\dots,d_{n-1}$};
\end{tikzpicture}}
\caption{\label{fig:glossary}
Tensor-train diagrams used in this study.  A circle is a
core, a vertical line is its physical index, and a horizontal line between
two cores is a bond; a line connecting two circles is summed over.  The
position of a core along the chain is its site.}
\end{figure}

Throughout this section we use the diagrammatic language of tensor networks,
which is summarized in Fig.~\ref{fig:glossary}.  A TT is a
product of three-index tensors $W^{(n)}\in\Rr^{r_{n-1}\times d_n\times r_n}$,
which we call the \emph{cores}.  The position $n$ of a core along the chain
(the horizontal direction in Fig.~\ref{fig:glossary}) is its \emph{site};
the variable $x_n$ attached to the core, drawn as a vertical line, is its
\emph{physical index}; and the line joining two neighboring cores is a
\emph{bond}, which carries a \emph{bond index} $a_n$ that is summed over.
The number $r_n$ of values of the bond index is the \emph{bond dimension}
(also called the TT rank).  Fixing the physical index of a core gives an
$r_{n-1}\times r_n$ matrix, which we write $W^{(n)}[x_n]$; this bracket
notation is used for all cores.

We use TT and core
rather than matrix product state and site tensor, because the object
represented is a classical distribution rather than a quantum state;
otherwise we use the standard vocabulary of the density-matrix
renormalization group~\cite{Schollwoeck2011}, in which the contraction of all
cores on one side of a bond is an \emph{environment} and a pass of local
updates over the sites is a \emph{sweep}.

\subsection{Settings}
\label{sec:setting}

We consider $N$ discrete variables $x_n\in\{1,\dots,d_n\}$ arranged on a
chain, and a bipartition of the chain into a left block $A=\{1,\dots,k\}$ and
a right block $B=\{k+1,\dots,N\}$, so that $\vx=(x_1,\dots,x_k)$ and
$\vy=(x_{k+1},\dots,x_N)$.  The joint distribution is given in the TT format,
\begin{equation}
  P(x_1,\dots,x_N)=W^{(1)}[x_1]\,W^{(2)}[x_2]\cdots W^{(N)}[x_N],
  \label{eq:Ptt}
\end{equation}
with cores $W^{(n)}[x_n]\in\Rr^{r_{n-1}\times r_n}$, $r_0=r_N=1$; we write
$r=\max_n r_n$ for the bond dimension of $P$.
The distribution is assumed normalized, $\sum_{x}P(x)=1$; if only an
unnormalized weight is available (e.g., a Boltzmann weight), the normalization
constant is obtained by one full contraction of the TT and is divided into any
one core.
In this study, the transformations are sought in the TT format,
\begin{align}
  f(\vx)&=F^{(1)}[x_1]\cdots F^{(k)}[x_k],
  \label{eq:ftt}\\
  g(\vy)&=G^{(k+1)}[x_{k+1}]\cdots G^{(N)}[x_N],
  \label{eq:gtt}
\end{align}
with cores $F^{(i)}[x_i],\,G^{(i)}[x_i]\in\Rr^{D_{i-1}\times D_i}$,
$D_0=D_k=D_N=1$, and a common bond dimension $D=\max_i D_i$,
which is the single variational parameter of the method.
For $D=1$, the
transformations are product functions $f(\vx)=\prod_i\phi_i(x_i)$; the linear
combinations of single-site functions $\sum_i\phi_i(x_i)$ are contained in
$D=2$.
Since $(f,g)$
satisfies the constraints exactly and the objective of Eq.~\eqref{eq:hgr}
is evaluated exactly, every pair produced by the algorithm is a feasible
point of the variational problem, i.e., it satisfies the constraints, and
the value of the objective is a lower bound to $\rhoHGR$.

\subsection{Outline of the algorithm}
\label{sec:outline}

The algorithm is an alternating maximization of the bilinear objective
\begin{equation}
  J(f,g)=\sum_{\vx,\vy}P(\vx,\vy)f(\vx)g(\vy)
  \label{eq:J}
\end{equation}
under the constraints of Eqs.~\eqref{eq:const1} and \eqref{eq:const2}.  
Bilinearity makes the alternation between $f$ and $g$ natural: for fixed $g$ the
objective is linear in $f$, so each update is a linear functional maximized
under one linear and one quadratic constraint, and the local problem discussed in
Sec.~\ref{sec:local} inherits this structure.
\begin{enumerate}
  \item Represent $P$ in the TT format as in Eq.~\eqref{eq:Ptt}, and initialize the
        cores of $f$ and $g$ (Sec.~\ref{sec:monotone}).
  \item Repeat until $J$ stops increasing (outer loop):
  \begin{enumerate}
    \item ($f$-update) with $g$ fixed, maximize $J$ over $f$ by a sweep of
          local core updates (Secs.~\ref{sec:reduction} and \ref{sec:local});
    \item ($g$-update) with $f$ fixed, maximize $J$ over $g$ in the same way.
  \end{enumerate}
  \item Evaluate $\rhoHGR\simeq J(f,g)$ by one full contraction
        (Sec.~\ref{sec:rho}).
\end{enumerate}
Each update (a) or (b) is analogous to the conditional-expectation
update of Sec.~\ref{sec:fixedpoint}, and inherits the structure of
Eqs.~\eqref{eq:gopt}--\eqref{eq:composed}; within an update, the sweep
over the cores is the alternation of alternating least squares (ALS)
\cite{Holtz2012}.  The word ``alternating'' thus appears at two levels,
between $f$ and $g$ as in ACE and between the cores of one function as in
ALS. I n this study, we call the former an \emph{update} and the latter a \emph{sweep}.
We now describe the $f$-update (a). The
$g$-update (b) is identical after the roles of the blocks $A$ and $B$ are exchanged, which in practice
means reversing the order of the sites so that the $B$-block is processed
from its right end.

\subsection{Reduction to a single core}
\label{sec:reduction}

With $g$ fixed, the objective in Eq.~\eqref{eq:J} depends on $f$ only through the
function
\begin{equation}
  \begin{split}
  b(\vx)&=\sum_{\vy}P(\vx,\vy)\,g(\vy),\\
  J(f,g)&=\sum_{\vx}f(\vx)\,b(\vx).
  \end{split}
  \label{eq:def_b}
\end{equation}
In the TT format, $b$ is obtained by contracting the physical indices of the
$B$-block of $P$ with the cores of $g$ as shown in Fig.~\ref{fig:tn}(b):
\begin{align}
  b(\vx)&=W^{(1)}[x_1]\cdots W^{(k)}[x_k]\,R_b,
  \label{eq:btt}\\
  R_b&=\sum_{\vy}\Bigl(W^{(k+1)}[x_{k+1}]\cdots W^{(N)}[x_N]\Bigr)
  \nonumber\\
  &\qquad\quad\times\Bigl(G^{(k+1)}[x_{k+1}]\cdots G^{(N)}[x_N]\Bigr),
  \label{eq:Rb}
\end{align}
where the vector $R_b\in\Rr^{r_k}$ is the $B$-block of $P$ contracted with
$g$.  As shown in Fig.~\ref{fig:tn}(b), $R_b$ is evaluated from the right
end, one site at a time, by contracting the pair $W^{(i)}[x_i]$ and
$G^{(i)}[x_i]$ for $i=N,\dots,k+1$, at a total cost of
$O\bigl((N-k)\,d\,(r^2D+rD^2)\bigr)$.
Thus $b$ is a TT on the $A$-block with the
cores of $P$ and the right boundary vector $R_b$.
In exactly the same way, the marginal distribution $\PA(\vx)$ is obtained by contracting the $B$-block of $P$ with the
all-ones tensor [Fig.~\ref{fig:tn}(c)],
\begin{equation}
  \begin{split}
  \PA(\vx)&=W^{(1)}[x_1]\cdots W^{(k)}[x_k]\,R_{1},\\
  R_{1}&=\sum_{\vy}W^{(k+1)}[x_{k+1}]\cdots W^{(N)}[x_N] .
  \end{split}
  \label{eq:R1}
\end{equation}
Both $R_b$ and $R_1$ are computed once per $f$-update.

Now choose a site $i\in A$ and consider its core $\beta:=F^{(i)}\in
\Rr^{D_{i-1}\times d_i\times D_i}$, keeping all other cores of $f$ fixed.
Because $f$ is a product of matrices, it is linear in each core separately;
to make this explicit, we split the product at site $i$,
\begin{align}
  f(\vx)&=F_{<i}[\vx]\,\beta[x_i]\,F_{>i}[\vx],
  \label{eq:fEbeta}\\
  F_{<i}[\vx]&:=F^{(1)}[x_1]\cdots F^{(i-1)}[x_{i-1}],
  \nonumber\\
  F_{>i}[\vx]&:=F^{(i+1)}[x_{i+1}]\cdots F^{(k)}[x_k],
  \label{eq:partial}
\end{align}
where the row vector $F_{<i}[\vx]\in\Rr^{1\times D_{i-1}}$ and the column
vector $F_{>i}[\vx]\in\Rr^{D_i\times 1}$ depend only on the variables to the
left and to the right of site $i$, respectively.  Collecting the three indices
of $\beta$ into one, $f=E\beta$ with the \emph{frame matrix}~\cite{Cichockietal2017}
(the retraction operator of Ref.~\cite{Holtz2012})
\begin{equation}
  E_{\vx,(\alpha x_i'\alpha')}:=
  \bigl(F_{<i}[\vx]\bigr)_{\alpha}\;\delta_{x_ix_i'}\;
  \bigl(F_{>i}[\vx]\bigr)_{\alpha'},
  \label{eq:frame}
\end{equation}
which
is a linear map from the core space $\Rr^{D_{i-1}d_iD_i}$ to the function
space $\Rr^{d_1\cdots d_k}$.  The frame matrix $E$ is never formed
explicitly; it appears only through the contractions of the next subsection.
Substituting $f=E\beta$ into Eq.~\eqref{eq:def_b} and into the
constraints of Eqs.~\eqref{eq:const1} and \eqref{eq:const2} gives the local problem
\begin{align}
  \max_{\beta}\quad & c^{\T}\beta,
  \label{eq:local_obj}\\
  \text{s.t.}\quad & d^{\T}\beta=0,\qquad \beta^{\T}G\beta=1,
  \label{eq:local_const}
\end{align}
with the vectors $c$, $d$ and the matrix $G$ defined by
[Figs.~\ref{fig:tn}(e)--\ref{fig:tn}(g)]
\begin{equation}
  c=E^{\T}b,\qquad d=E^{\T}\PA,\qquad G=E^{\T}\diag(\PA)\,E,
  \label{eq:cdG}
\end{equation}
all of dimension $D_{i-1}d_iD_i$.  Explicitly,
$c^{\T}\beta=\sum_{\vx}f(\vx)b(\vx)=J(f,g)$, $d^{\T}\beta=\sum_{\vx}\PA(\vx)
f(\vx)=\E[f]$, and $\beta^{\T}G\beta=\sum_{\vx}\PA(\vx)f(\vx)^2=\E[f^2]$.  The
matrix $G$ is the Gram matrix of the frame matrix in the $\PA$-weighted inner
product, hence symmetric positive semidefinite; because of the
$\delta_{x_ix_i'}$ in Eq.~\eqref{eq:frame}, it is block diagonal in the
physical index $x_i$,
\begin{equation}
  \begin{split}
  G_{(\alpha x_i\alpha'),(\gamma x_i'\gamma')}&=\delta_{x_ix_i'}\,
  G^{[x_i]}_{(\alpha\alpha'),(\gamma\gamma')},\\
  G^{[x_i]}&\in\Rr^{D_{i-1}D_i\times D_{i-1}D_i}.
  \end{split}
  \label{eq:Gblock}
\end{equation}
The inverse of the local metric $G$ plays the role of the division by
$\PA$; in the algorithm, the normalization by the marginal, $D_A^{-1/2}$ in Eq.~\eqref{eq:K}, enters nowhere else.

\subsection{Closed-form solution of the local problem}
\label{sec:local}

The problem of Eqs.~\eqref{eq:local_obj} and \eqref{eq:local_const} maximizes a linear
functional over the intersection of an ellipsoid and a hyperplane and has a
unique solution whenever $G$ is positive definite on the relevant
subspace.  With Lagrange multipliers $\mu,\lambda$, the Lagrangian is
\begin{equation}
  \begin{split}
  \mathcal L&=c^{\T}\beta-\frac{\mu}{2}\bigl(\beta^{\T}G\beta-1\bigr)
  -\lambda\,d^{\T}\beta,\\
  \partial_\beta\mathcal L&=c-\mu G\beta-\lambda d=0,
  \end{split}
\end{equation}
so that $\beta\propto G^{-1}(c-\lambda d)$.  The multiplier $\lambda$ is
determined by the zero-mean constraint $d^{\T}\beta=0$, and $\mu$ by the
normalization:
\begin{equation}
  \beta=\frac{\tilde\beta}{\sqrt{\tilde\beta^{\T}G\tilde\beta}},\qquad
  \tilde\beta=G^{-1}(c-\lambda d),\qquad
  \lambda=\frac{d^{\T}G^{-1}c}{d^{\T}G^{-1}d} .
  \label{eq:betasol}
\end{equation}
The Lagrange condition only identifies a stationary point; that
Eq.~\eqref{eq:betasol} is the maximum, with the maximum value
$J_{\max}=\sqrt{(c-\lambda d)^{\T}G^{-1}(c-\lambda d)}$, is shown in
Appendix~\ref{app:local}, where the change of variables $\gamma=G^{1/2}\beta$
turns the problem into the projection of a vector onto a hyperplane.

Two remarks are in order.  (i) Because of Eq.~\eqref{eq:Gblock} the linear
system for $\tilde\beta$
decouples into $d_i$ independent linear systems of dimension $D_{i-1}D_i$,
one for each value of $x_i$, each with the two right-hand sides $c^{[x_i]}$ and
$d^{[x_i]}$; the scalar $\lambda$ couples the blocks only through the
inner products in Eq.~\eqref{eq:betasol}. 
(ii) In practice, we do not compute $G^{-1}$ directly. Instead, we solve the least-squares problem for each $x_i$, 
$\min_{\beta_1,\beta_2}\|G^{[x_i]}[\beta_1\;\beta_2]-[c^{[x_i]}\;d^{[x_i]}]\|$
(e.g., with a rank-revealing least-squares routine and a relative cutoff on
the singular values of $G^{[x_i]}$), add a ridge of order
$10^{-10}\,\mathrm{tr}\,G^{[x_i]}/(D_{i-1}D_i)$ for numerical stability, and
assemble $\tilde\beta=\beta_1-\lambda\beta_2$.
Null directions of $G$ correspond to
components of the frame matrix that have zero probability under $\PA$
(configurations outside the support); they do not affect $J$, and the
cutoff removes them from the update.  

\begin{figure*}
\centering
\begin{tikzpicture}[x=0.88cm,y=0.9cm]
  \begin{scope}[shift={(0,0)}]
    \node at (-1.6,0) {(a)};
    \Pchain{P}{0}
    \foreach \i in {0,...,4}{\draw[leg] (P\i)--++(0,-0.5);}
    \draw[brace] (-0.3,0.35)--(2.3,0.35) node[midway,above=4pt]{$\vX$};
    \draw[brace] (2.7,0.35)--(4.3,0.35) node[midway,above=4pt]{$\vY$};
    \node[right=0.4cm of P4] {$=P(\vx,\vy)$};
  \end{scope}
  \begin{scope}[shift={(0,-2.4)}]
    \node at (-1.6,0) {(b)};
    \Pchain{P}{0}
    \foreach \i in {0,1,2}{\draw[leg] (P\i)--++(0,-0.5);}
    \foreach \i in {3,4}{\node[core] (G\i) at (\i,-1) {};\draw[leg] (P\i)--(G\i);}
    \draw[bond] (G3)--(G4);
    \draw[mbrace] (2.7,-1.3)--(4.3,-1.3) node[midway,below=4pt]{$g(\vy)$};
    \node[right=0.4cm of P4] {$=b(\vx)$};
  \end{scope}
  \begin{scope}[shift={(0,-5.2)}]
    \node at (-1.6,0) {(c)};
    \Pchain{P}{0}
    \foreach \i in {0,1,2}{\draw[leg] (P\i)--++(0,-0.5);}
    \foreach \i in {3,4}{\node[ones] (O\i) at (\i,-1) {$1$};\draw[leg] (P\i)--(O\i);}
    \draw[bond] (O3)--(O4);
    \draw[mbrace] (2.7,-1.3)--(4.3,-1.3) node[midway,below=4pt]{$\mathbf 1$};
    \node[right=0.4cm of P4] {$=\PA(\vx)$};
  \end{scope}
  \begin{scope}[shift={(0,-8.0)}]
    \node at (-1.6,0) {(d)};
    \Pchain{P}{0}
    \node[core] (F0) at (0,-1) {};
    \node[beta] (F1) at (1,-1) {};
    \node[core] (F2) at (2,-1) {};
    \node[core] (G3) at (3,-1) {};
    \node[core] (G4) at (4,-1) {};
    \draw[bond] (F0)--(F1)--(F2);
    \draw[bond] (G3)--(G4);
    \foreach \i in {0,1,2}{\draw[leg] (P\i)--(F\i);}
    \foreach \i in {3,4}{\draw[leg] (P\i)--(G\i);}
    \node[below=3pt of F1,text=orange!70!black] {$\beta$};
    \draw[mbrace] (-0.3,-1.3)--(0.3,-1.3) node[midway,below=4pt]{$F_{<i}$};
    \draw[mbrace] (1.7,-1.3)--(2.3,-1.3) node[midway,below=4pt]{$F_{>i}$};
    \draw[mbrace] (2.7,-1.3)--(4.3,-1.3) node[midway,below=4pt]{$g(\vy)$};
    \node[right=0.4cm of P4] {$=J(f,g)$};
  \end{scope}
  \begin{scope}[shift={(9.1,0)}]
  \begin{scope}[shift={(0,0)}]
    \node at (-1.6,-0.5) {(e)};
    \Pchain{P}{0}
    \node[core] (F0) at (0,-1) {};
    \node[hole] (H1) at (1,-1) {};
    \node[core] (F2) at (2,-1) {};
    \node[core] (G3) at (3,-1) {};
    \node[core] (G4) at (4,-1) {};
    \draw[bond] (F0)--(H1)--(F2);
    \draw[bond] (G3)--(G4);
    \foreach \i in {0,2}{\draw[leg] (P\i)--(F\i);}
    \draw[leg] (P1)--(H1);
    \foreach \i in {3,4}{\draw[leg] (P\i)--(G\i);}
    \draw[mbrace] (2.7,-1.3)--(4.3,-1.3) node[midway,below=4pt]{$g(\vy)$};
    \node[right=0.4cm of P4] {$=c_{\alpha x_2\alpha'}$};
  \end{scope}
  \begin{scope}[shift={(0,-2.9)}]
    \node at (-1.6,-0.5) {(f)};
    \Pchain{P}{0}
    \node[core] (F0) at (0,-1) {};
    \node[hole] (H1) at (1,-1) {};
    \node[core] (F2) at (2,-1) {};
    \node[ones] (O3) at (3,-1) {$1$};
    \node[ones] (O4) at (4,-1) {$1$};
    \draw[bond] (F0)--(H1)--(F2);
    \draw[bond] (O3)--(O4);
    \foreach \i in {0,2}{\draw[leg] (P\i)--(F\i);}
    \draw[leg] (P1)--(H1);
    \foreach \i in {3,4}{\draw[leg] (P\i)--(O\i);}
    \draw[mbrace] (2.7,-1.3)--(4.3,-1.3) node[midway,below=4pt]{$\mathbf 1$};
    \node[right=0.4cm of P4] {$=d_{\alpha x_2\alpha'}$};
  \end{scope}
  \begin{scope}[shift={(0,-7.0)}]
    \node at (-1.6,0) {(g)};
    \Pchain{P}{0}
    \node[core] (F0) at (0,-1) {};
    \node[hole] (H1) at (1,-1) {};
    \node[core] (F2) at (2,-1) {};
    \draw[bond] (F0)--(H1)--(F2);
    \foreach \i in {0,2}{\draw[leg] (P\i)--(F\i);}
    \draw[leg] (P1)--(H1);
    \node[core] (U0) at (0,1) {};
    \node[hole] (K1) at (1,1) {};
    \node[core] (U2) at (2,1) {};
    \draw[bond] (U0)--(K1)--(U2);
    \foreach \i in {0,2}{\draw[leg] (P\i)--(U\i);}
    \draw[leg] (P1)--(K1);
    \node[ones] (O3) at (3,-1) {$1$};
    \node[ones] (O4) at (4,-1) {$1$};
    \draw[bond] (O3)--(O4);
    \foreach \i in {3,4}{\draw[leg] (P\i)--(O\i);}
    \draw[mbrace] (2.7,-1.3)--(4.3,-1.3) node[midway,below=4pt]{$\mathbf 1$};
    \draw[brace] (-0.3,1.35)--(2.3,1.35) node[midway,above=4pt]{$E$};
    \draw[mbrace] (-0.3,-1.3)--(2.3,-1.3) node[midway,below=4pt]{$E$};
    \node[right=0.4cm of P4] {$=G^{[x_2]}_{(\alpha\alpha'),(\gamma\gamma')}$};
  \end{scope}
  \end{scope}
\end{tikzpicture}
\caption{\label{fig:tn}
Tensor-network representation of the algorithm ($N=5$, $k=3$).  Shaded
circles are the cores of $P$, open circles those of $f$ and $g$. Horizontal
lines indicate bond indices, vertical lines physical indices, and a line joining
two cores is summed over.  (a) the joint
distribution $P$ in Eq.~\eqref{eq:Ptt}.
(b) the $B$-block of $P$ contracted with $g$ as in Eq.~\eqref{eq:def_b}. 
(c) the marginal $\PA$, the $B$-block
contracted with the all-ones tensor, As Eq.~\eqref{eq:R1}. 
(d) the objective $J(f,g)$.  The orange core $\beta$ at site $i=2$ is the one being optimized;
the cores of $f$ to its left and right form $F_{<i}$ and $F_{>i}$ of
Eq.~\eqref{eq:partial}, i.e., the frame matrix $E$ of Eq.~\eqref{eq:frame}.
Right column, the three ingredients of the local problem at site $i=2$,
Eq.~\eqref{eq:cdG}.  The dashed circle marks the position of $\beta$. its
three indices are left open, so that each network has the shape of $\beta$.
(e) $c=E^{\T}b$, the network in panel (d) without $\beta$. (f) $d=E^{\T}\PA$, the
same with $g$ replaced by the all-ones tensor, so that $d^{\T}\beta=\E[f]$;
(g) $G=E^{\T}\diag(\PA)E$, two copies of $E$ attached to the same physical
indices of $P$, which implements $\diag(\PA)$, so that
$\beta^{\T}G\beta=\E[f^2]$.  The physical index of $P$ at site $2$ is shared
by both holes and stays open, which is why $G$ is block diagonal in $x_2$,
Eq.~\eqref{eq:Gblock}.  The left and right parts of each network, cut at
site $2$, are the environments of Sec.~\ref{sec:local}.}
\end{figure*}
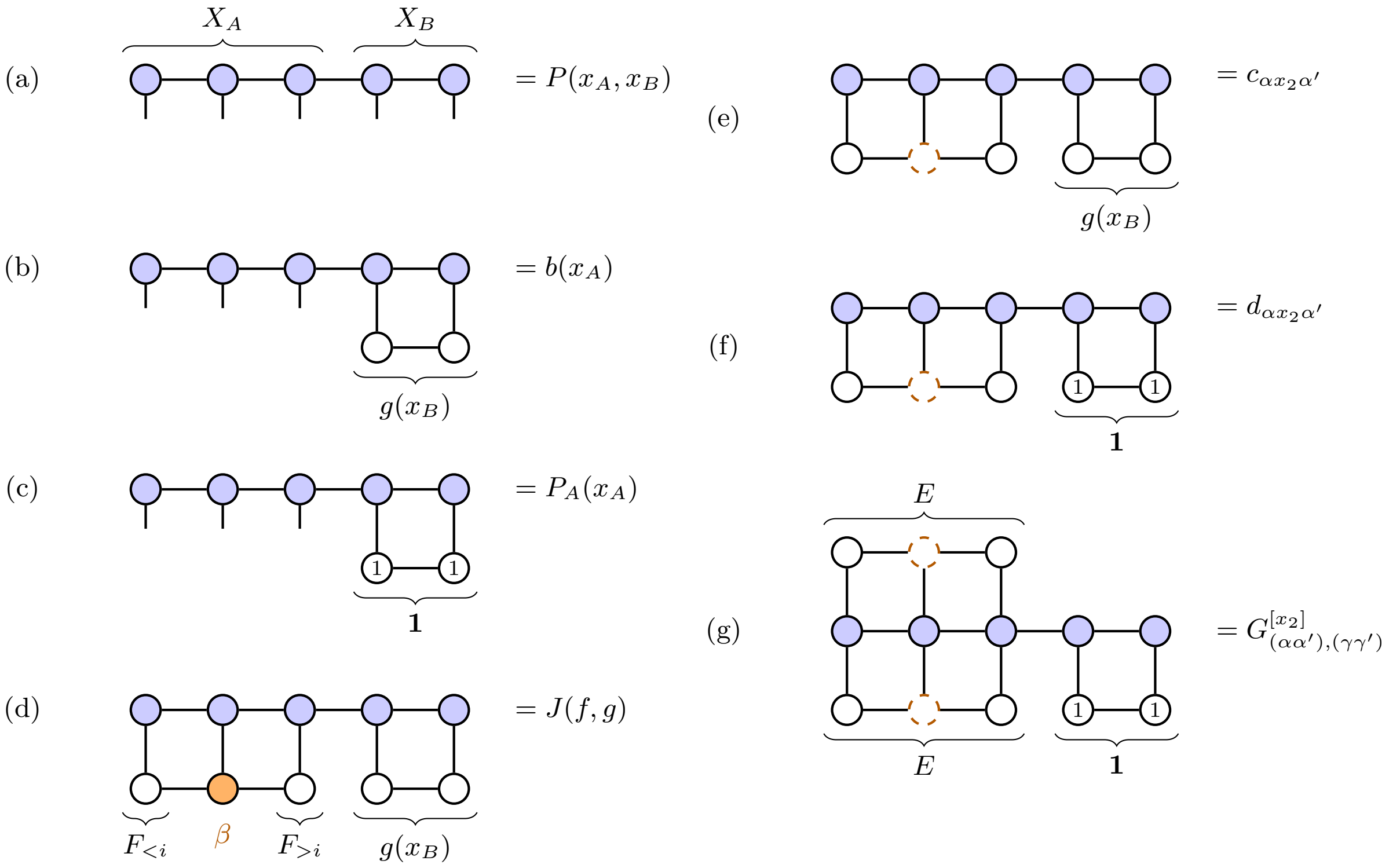

Unlike the standard alternating linear scheme for TT approximation
\cite{Holtz2012}, where an orthogonal (canonical) gauge of the non-optimized
cores makes the local metric the identity, the metric here is weighted by
$\PA$ and cannot be gauged away; gauging the environment cores merely improves
the conditioning of $G$.  A local metric that is not the identity is
not peculiar to this problem: it appears in the TT formulation of (linear)
canonical correlation analysis \cite{Cichocki2014,Cichockietal2017}, where it
is the Gram matrix of a data matrix seen through the frame matrix, and in the
TT computation of eigenfunctions of transfer operators
\cite{Nuske2021}, where it is the Gram matrix of the basis functions
evaluated on the data.  What differs here is where the metric comes from:
$G$ is contracted from the distribution $P$ itself, Fig.~\ref{fig:tn}(g), so
that no data matrix is formed and the weighting by $\PA$ is exact.

The quantities $c$, $d$, and $G$ are not computed from scratch at every
site.  Each of the three networks in Figs.~\ref{fig:tn}(e)--\ref{fig:tn}(g) is cut at the
site being updated into a left and a right part, i.e., the environments. After
each local update, the environment on one side is extended by one site, and a
sweep over the $k$ cores of $f$ is a right-to-left pass followed by a
left-to-right pass, as in density matrix renormalization group~\cite{Schollwoeck2011}.  One sweep over the
$k$ cores of $f$ costs $O\bigl(k\,(dD^6+r\,dD^4+r^2dD^2)\bigr)$ operations
(dense least-squares solves, assembly of the blocks $G^{[x_i]}$, and
extension of the environments, respectively), and the $g$-update is
analogous with $k\to N-k$: the cost is linear in the system size and
polynomial in the two bond dimensions, and at no stage is a tensor with more
than three physical indices formed.

\subsection{Evaluation of $\rhoHGR$ and of the transformations}
\label{sec:rho}

Once the alternation has converged, the HGR maximal correlation is obtained from the
full contraction of Fig.~\ref{fig:tn}(d),
\begin{multline}
  \rhoHGR\simeq J(f,g)=\sum_{\vx,\vy}P(\vx,\vy)f(\vx)g(\vy)\\
  =\Bigl(\sum_{\vx}W[\vx]\,F[\vx]\Bigr)\cdot
   \Bigl(\sum_{\vy}W[\vy]\,G[\vy]\Bigr),
  \label{eq:rhoTT}
\end{multline}
at cost of $O\bigl(N\,d\,(r^2D+rD^2)\bigr)$, where the two factors are the
contractions of the
$A$- and $B$-blocks, analogous to $R_b$, meeting at bond $k$.  The optimal
transformations are available as TTs in Eqs.~\eqref{eq:ftt}--\eqref{eq:gtt},
and can be evaluated at any configuration by $k$ (or $N-k$) matrix-vector
products, so that $f^{*}$ and $g^{*}$ can be inspected or compared with
candidate order parameters without expanding them on the full configuration
space.

\subsection{Convergence, initialization, and restarts}
\label{sec:monotone}

Each local update keeps $f$ feasible, because $d^{\T}\beta=\E[f]$ and
$\beta^{\T}G\beta=\E[f^2]$ are the exact constraints on $f$ given the other
cores, and it can only increase $J$, because the previous core is itself a
feasible point of Eqs.~\eqref{eq:local_obj} and \eqref{eq:local_const}.  The same
holds for the $g$-sweeps.  Hence the sequence of objective values generated by
the algorithm is nondecreasing and bounded above by $\rhoHGR$, and it
converges.  The increase of $J$ over one alternation between $f$ and $g$
can serve as a stopping criterion; in the computations reported below,
each update consists of a single sweep over the cores of its block, and the
number of alternations is fixed in advance (Secs.~\ref{sec:synth_results}
and~\ref{sec:afm_results}).
Convergence of the value to a local maximum of the TT-restricted problem,
rather than to $\rhoHGR$ itself, cannot be excluded in general; we
address this empirically by the comparisons with exact references in
Secs.~\ref{sec:synthetic} and~\ref{sec:afm} and
Appendix~\ref{app:svdcheck}, and by the dependence on $D$ and on the
initial point reported at the end of Sec.~\ref{sec:afm_results}.

In practice, the cores of $f$ and $g$ are initialized as the all-ones tensor
plus Gaussian noise (standard deviation $0.3$ for random restarts, $0.05$ for
a warm start); since the all-ones tensor represents the constant function,
the first local solve immediately projects out the trivial mode through the
constraint $d^{\T}\beta=0$.  When a natural single-site profile $h(x)$ is
known (e.g., $\cos\theta$ for angular variables), the first run is
warm-started by initializing the core at site $k$, the site adjacent to the
cut, as $h(x_k)$ times the all-ones tensor plus the smaller noise.
Several random restarts are run and the largest value of $J$ is
kept; because every value is a lower bound to $\rhoHGR$, the maximum over
restarts is the tightest certified bound available.

\section{Nonlinear synthetic distributions}
\label{sec:synthetic}

The exact SVD of Sec.~\ref{sec:svd} requires the full table of $P$ and
is therefore limited to systems of a few variables.  In this section, we
construct distributions of many variables whose optimal transformations
are known in closed form, so that the algorithm can be tested at sizes
where $P$ cannot be written as a table.  The
construction also shows what the algorithm does with a distribution that
hides a nonlinear relation between the two blocks: it returns the
relation.  Section~\ref{sec:synth_constr} gives the construction and
the closed-form reference values, Sec.~\ref{sec:synth_models} the four
distributions used, and Sec.~\ref{sec:synth_results} the results.
 
\subsection{Construction}
\label{sec:synth_constr}

We construct a distribution in which the dependence between the two
blocks is carried entirely by a prescribed pair of statistics, $S_X(\vx)$
and $S_Y(\vy)$, which are tied to each other up to noise, $S_X\simeq
S_Y$.  Since $P$ then depends on $\vx$ only through $S_X$ and on $\vy$
only through $S_Y$, the optimal transformations are functions of $S_X$
and $S_Y$ (Appendix~\ref{app:suff}), and the algorithm can be asked
whether it finds them.  We take the two statistics to be sums of
functions of single variables,
\begin{equation}
  S_X=\sum_{n\le k}\epsilon_n\,a_n(x_n),\qquad
  S_Y=-\sum_{n>k}\epsilon_n\,a_n(x_n),
  \label{eq:SXSY}
\end{equation}
where $a_n$ is a fixed function of the single variable $x_n$ (for
example $a_n(x)=x$, $x^2/2$ or $\sin x$; a variable that does not take
part has $a_n\equiv0$) and $\epsilon_n=\pm1$ is a sign.  The sum form is what
makes the TT representation below possible; the functions $a_n$ are
otherwise free, so the relation between the two blocks may be
nonlinear.

The simplest way to tie $S_X$ to $S_Y$ would be a Gaussian weight in
$S_X-S_Y$, but a Gaussian of a sum of many variables has no useful TT
format.  We therefore take all variables to be angles, $x_n\in\{-\pi+2\pi
(j+\tfrac12)/d,\;j=0,\dots,d-1\}$ with a uniform prior, read $S_X$ and
$S_Y$ modulo $2\pi$, and tie them by a wrapped Gaussian written as a
Fourier series,
\begin{equation}
  P(x_1,\dots,x_N)\propto C(S_X-S_Y)
  =C\Bigl(\sum_{n=1}^{N}\epsilon_n\,a_n(x_n)\Bigr),
  \label{eq:synth}
\end{equation}
\begin{equation}
  C(u)=\sum_{m=-\infty}^{\infty}c_m\,e^{imu},\qquad c_m=e^{-\sigma^2m^2/2}.
  \label{eq:kernel}
\end{equation}
This choice of $C$ has three useful properties.  First, by Poisson
summation the series equals the Gaussian of width $\sigma$ wrapped
around the circle,
\begin{equation}
  C(u)=\frac{\sqrt{2\pi}}{\sigma}\sum_{l=-\infty}^{\infty}
  e^{-(u-2\pi l)^2/2\sigma^2},
  \label{eq:wrapped}
\end{equation}
which, up to the constant factor, is the wrapped normal distribution of
directional statistics; so Eq.~\eqref{eq:synth}
says that $S_X$ equals $S_Y$ up to Gaussian noise of width $\sigma$,
the noise being wrapped around the circle, i.e., the distribution
encodes the relation $S_X\simeq S_Y\pmod{2\pi}$ and nothing else.
Second, in the Fourier form each mode of the sum factorizes into a
product over the variables, $e^{im\sum_ns_na_n(x_n)}=\prod_n
e^{ims_na_n(x_n)}$, so that $C(S_X-S_Y)$ is a sum of product functions,
one per mode $m$, and a sum of product functions is a TT whose bond
dimension is the number of modes.  Appendix~\ref{app:rotation} gives
the cores explicitly, in a real form in which the bond carries the
constant mode and the pair $(\cos mS,\sin mS)$ for each $m\ge1$, $S$
being the partial sum up to the current site, and each core rotates the
pair by the angle $m\,s_na_n(x_n)$.  Third, the coefficients $c_m$
decay as a Gaussian in $m$, so the series can be truncated at an order
$J$ with an error controlled by the tail,
\begin{equation}
  \Bigl|C(u)-\sum_{m=-M_c}^{M_c}c_m\,e^{imu}\Bigr|\le 2\sum_{m>M_c}c_m
  \qquad\text{for all }u ,
  \label{eq:trunc}
\end{equation}
and we choose $M_c$ such that this bound is below $10^{-13}$; the
truncated kernel is then still strictly positive (its minimum,
$C(\pi)$, is of order $10^{-3}$ for the value of $\sigma$ used below),
and the distribution is an exact TT of bond dimension $r=2M_c+1$.  The
construction works for any choice of the $a_n$ and any number of
variables, so the distribution can contain nonlinear functions of
single angles and can be made larger than any table; and because the
$c_m$ are known in closed form, the exact singular values of the
problem follow at once, as shown next.
 
By Appendix~\ref{app:suff}, the optimal
transformations are functions of the two sums, $f^{*}=h_f(S_X)$
and $g^{*}=h_g(S_Y)$, because $P$ depends on $\vx$ only through $S_X$
and on $\vy$ only through $S_Y$.  The functions $h_f$, $h_g$ and the
value of $\rhoHGR$ follow from the Fourier form of the kernel.  Suppose
that at least one variable of block $A$ enters the sum as $\pm x_n$,
i.e., with $a_n(x)=x$.  Since $x_n$ is uniform on the grid,
$\E[e^{imx_n}]=0$ for $0<|m|<d$, and the same then holds for $S_X$
whatever the other terms of the sum are: $\E[e^{imS_X}]=0$ for
$0<|m|<d$.  If the same holds for block $B$, the marginals $\PA$ and
$\PB$ are uniform, and the functions $e^{imS_X}$ and $e^{imS_Y}$ with
$|m|\le J<d/2$ are orthonormal under $\PA$ and $\PB$, respectively.  Then
$P\propto C(S_X-S_Y)=\sum_mc_m\,e^{imS_X}\,e^{-imS_Y}$ is already an SVD
of the normalized matrix $K$ of Eq.~\eqref{eq:K} (in real form, with
$\cos mS$ and $\sin mS$ in place of $e^{\pm imS}$), and its singular
values are the Fourier coefficients of the kernel,
\begin{equation}
  \sigma_1=1,\qquad \sigma_{2m}=\sigma_{2m+1}=e^{-\sigma^2m^2/2}
  \quad(m=1,2,\dots),
  \label{eq:synth_spectrum}
\end{equation}
each $m\ge1$ appearing twice, for $\cos mS$ and $\sin mS$.  Hence
\begin{equation}
  \begin{split}
  &\rhoHGR=e^{-\sigma^2/2},\\
  &f^{*}=\sqrt2\cos(S_X-\alpha),\qquad g^{*}=\sqrt2\cos(S_Y-\alpha),
  \end{split}
  \label{eq:synth_rho}
\end{equation}
with the same, arbitrary phase $\alpha$ in $f^{*}$ and $g^{*}$; the
factor $\sqrt2$ makes $\E[f^{*2}]=1$.  The optimal transformation is
thus a two-dimensional space, and any $\alpha$ is equally good; this is
the degenerate case of Eq.~\eqref{eq:kappa_deg}.  If no variable of a
block enters linearly, $S_X$ is not uniform in the above sense,
Eq.~\eqref{eq:synth_rho} does not apply, and the reference must be
taken from the exact SVD of the full table (model 3 below).  
In either case, the algorithm is given neither $S_X$ nor $S_Y$; the comparison
tests whether it finds the functions $h_f(S_X)$ and $h_g(S_Y)$ on its
own.
 
\subsection{Models}
\label{sec:synth_models}
 
We use four distributions of the form in Eq.~\eqref{eq:synth}, listed in
Table~\ref{tab:synth}, all with $d=32$, $\sigma=0.8$ and $J=9$, at
which the truncation bound of Eq.~\eqref{eq:trunc} is
$2.5\times10^{-14}$ and $C(\pi)=2.8\times10^{-3}$.  With these values
the closed-form reference
of Eq.~\eqref{eq:synth_rho} is $\rhoHGR=e^{-\sigma^2/2}=0.726149\ldots$,
the TT of Eq.~\eqref{eq:synth} has bond dimension $2J+1=19$, and the
next singular value is $\sigma_4=e^{-2\sigma^2}=0.278037$, which agrees
with the exact SVD of the full table (Appendix~\ref{app:svdcheck}).
\begin{table*}
\caption{\label{tab:synth}
The four synthetic distributions.  The chain order is the order in
which the variables are placed on the sites of the TT; the cut is after
the first $k$ sites.  In model~2 the variables are the same as in
model~1, but the TT is written in a different order, so that the cut
separates a different pair of blocks.  Model~3, with
$\varphi(x_1,x_3)=x_1x_3/2\pi+x_1^4/2\pi^3$, is not of the additive
form~\eqref{eq:synth} and is built as a table and compressed to a TT by
successive SVDs (bond dimension 25, exact to $10^{-12}$); the other
three are exact TTs of bond dimension 19 by the construction of
Appendix~\ref{app:rotation}.
$S_X$ and $S_Y$ are the statistics that the optimal transformations
depend on.}
\begin{ruledtabular}
\begin{tabular}{lllll}
 & chain order (cut $|$) & $S_X$ & $S_Y$ & reference $\rhoHGR$ \\
\hline
Model 1 (linear, split A) & $x_1\,x_2\,|\,x_3\,x_4$ & $x_1+x_2$ & $x_3+x_4$ & $e^{-\sigma^2/2}$, exact SVD \\
Model 2 (linear, split B) & $x_2\,x_4\,|\,x_1\,x_3$ & $x_2-x_4$ & $x_3-x_1$ & $e^{-\sigma^2/2}$, exact SVD \\
Model 3 (cross term)      & $x_1\,x_2\,x_3\,|\,x_4$ & $\varphi(x_1,x_3)$ & $x_4$ & exact SVD \\
Model 4 (ten variables)   & $x_1\cdots x_8\,|\,x_9\,x_{10}$ & $\sin x_1+x_2^2/2+x_3$ & $x_9+x_{10}$ & $e^{-\sigma^2/2}$ \\
\end{tabular}
\end{ruledtabular}
\end{table*}
 
\emph{Models 1 and 2} are the same distribution,
$P\propto C(x_1+x_2-x_3-x_4)$, which encodes the linear relation
$x_1+x_2\simeq x_3+x_4$.  In model~1 the blocks are $(x_1,x_2)$ and
$(x_3,x_4)$.  In model~2 the same distribution is written in the chain
order $(x_2,x_4,x_1,x_3)$ and cut after two sites, so that the blocks are
$(x_2,x_4)$ and $(x_1,x_3)$.  The relation between these two blocks is
$x_2-x_4\simeq x_3-x_1$, which is the same relation rearranged.  The two
models test whether the algorithm finds the relation that belongs to
the chosen bipartition, rather than a function determined by the
distribution irrespective of where it is cut.
 
\emph{Model 3} adds a cross term.  With $\varphi(x_1,x_3)=x_1x_3/2\pi+x_1^4/2\pi^3$
(the coefficients keep $\varphi$ inside one period, so that no wrapping
occurs), the distribution is $P\propto C(\varphi(x_1,x_3)-x_4)$ and $x_2$
is independent of the rest.  The blocks are $(x_1,x_2,x_3)$ and $(x_4)$.
This model is not of the additive form~\eqref{eq:synth}, because of the
product $x_1x_3$; its TT is obtained from the table by successive SVDs
and has bond dimension 25.  Here $S_X=\varphi(x_1,x_3)$ is not uniform,
so Eq.~\eqref{eq:synth_rho} does not apply, and the reference is the
exact SVD of the full table ($32^4\simeq10^6$ entries),
$\sigma_2=0.551890$; the spectrum is not degenerate
($\sigma_3=0.319734$).  The model tests two things: that
the recovered $f$ is a function of a nonlinear, non-additive statistic,
and that it ignores the irrelevant variable $x_2$.
 
\emph{Model 4} has ten variables, $P\propto C(\sin x_1+x_2^2/2+x_3-x_9-x_{10})$,
with the cut after $x_8$; the five variables $x_4,\dots,x_8$ do not enter
the sum.  The configuration space has $32^{10}\simeq1.1\times10^{15}$
points, so neither $P$ nor $f$ can be written as a table, but the TT of
$P$ has bond dimension 19 and the reference value is the closed form
$e^{-\sigma^2/2}$: although $\sin x_1+x_2^2/2$ by itself is not
uniform, $x_3$ enters $S_X$ as $+x_3$, which makes $S_X$ uniform in the
sense of Sec.~\ref{sec:synth_constr}, and $x_9+x_{10}$ is uniform as
well.
The model tests
the algorithm at a size that is out of reach of the exact SVD, with a
nonlinear statistic ($\sin x_1$ and $x_2^2$) and with irrelevant
variables.
 
\subsection{Results}
\label{sec:synth_results}
 
The algorithm was run with $D=6$ (models 1, 2, 4) and $D=8$ (model 3),
with six to ten $f$- and $g$- updates.  The first run was warm-started from
$f_0=\cos S_X$ and $g_0=\cos S_Y$ (an exact rank-2 TT, padded with noise
to bond dimension $D$); a random start as in Sec.~\ref{sec:monotone} gives
the same values of $\rho_{\rm TT}$ within $10^{-9}$ for models 1 and 4
(four and three random seeds, respectively), so the warm start only
fixes the phase $\alpha$ of Eq.~\eqref{eq:synth_rho}.  The run times are
about 1 s for models 1 and 2, 20 s for model 3 and 8 s for model
4, on a laptop.
 
\emph{Value of the correlation.}  For models 1, 2, and 4 the algorithm
returns $\rho_{\rm TT}=0.726149$, equal to the closed form
$e^{-\sigma^2/2}$ to $10^{-15}$; for model 3 it returns
$\rho_{\rm TT}=0.551890$, equal to $\sigma_2$ to $4\times10^{-8}$.  In
all cases $\rho_{\rm TT}\le\rhoHGR$.  For models 1--3, the full table
of $P$ can still be built and the exact SVD of $K$ computed; the
comparison, given in Appendix~\ref{app:svdcheck}
(Table~\ref{tab:svd}), shows that $\rho_{\rm TT}$ agrees with
$\sigma_2$ to between $10^{-16}$ and $10^{-8}$ and the recovered $f$
and $g$ with the exact singular vectors to six digits.  For model 4
the reference cannot be checked by any other method than the closed
form.
 
\begin{figure*}
\includegraphics[width=\textwidth]{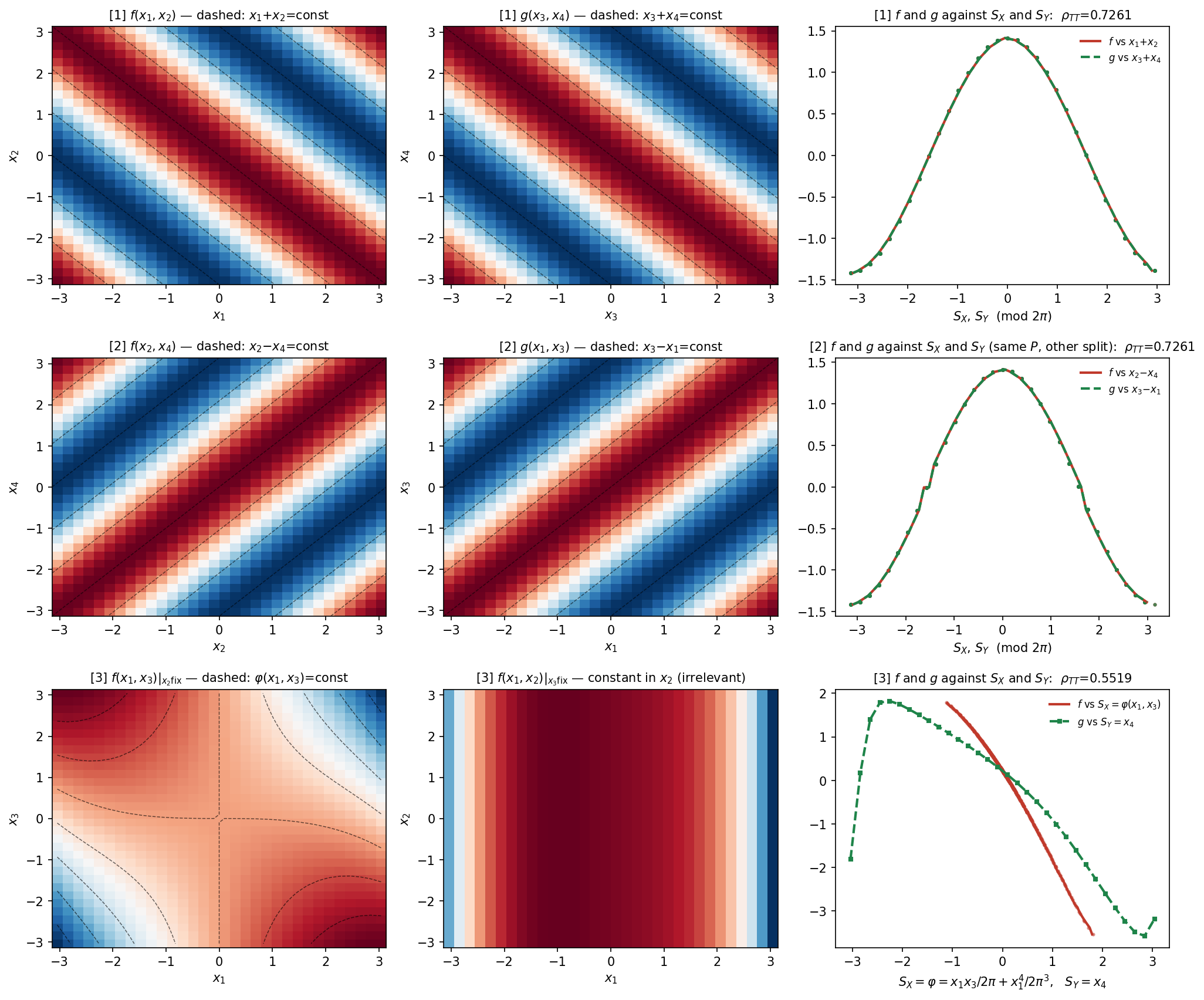}
\caption{\label{fig:synth_small}
Recovered optimal transformations for the synthetic models 1--3
(one model per row).  Left and middle: $f$ and $g$ as heat maps on the
grid; black dashed lines are the level sets of the statistics $S_X$ and $S_Y$
of Table~\ref{tab:synth}.  For model 3 the left panel is a slice of
$f(x_1,x_2,x_3)$ at fixed $x_2$ with the level sets of
$\varphi(x_1,x_3)$, and the middle panel a slice at fixed $x_3$, which is
constant along the irrelevant variable $x_2$.  Right: the values of $f$
(red) and $g$ (green) plotted against the values of $S_X$ and $S_Y$ for
all grid points; the collapse onto a single curve shows that $f$ and $g$
are functions of the statistics.  For model 3, $g(x_4)$ is shown on the
right axis.  The values $\rho_{\rm TT}$ are given in the titles.}
\end{figure*}

\begin{figure*}
\includegraphics[width=\textwidth]{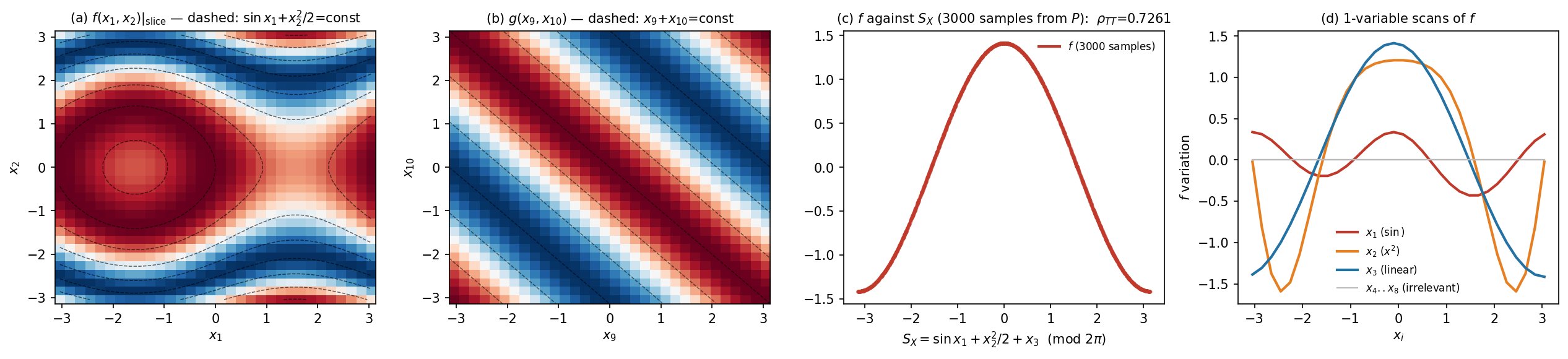}
\caption{\label{fig:synth_big}
Model 4 (ten variables, $32^{10}$ configurations).  (a) Slice of $f$
in the $(x_1,x_2)$ plane with the other variables fixed; black dashed lines
are the level sets of $\sin x_1+x_2^2/2$.  (b) $g(x_9,x_{10})$ with the
level sets of $x_9+x_{10}$.  (c) Values of $f$ at 3000 configurations
sampled from $P$ against $S_X=\sin x_1+x_2^2/2+x_3$ (mod $2\pi$).  (d)
Change of $f$ when a single variable is varied with the others fixed:
the three relevant variables (colored) and the five irrelevant ones
(gray).}
\end{figure*}

\emph{Form of the transformations.}  Figure~\ref{fig:synth_small} shows
the recovered $f$ and $g$ for models 1--3 and
Fig.~\ref{fig:synth_big} for model 4.  In each heat map, the color is
the value of the recovered transformation on a two-dimensional slice of
the configuration space, and the black dashed lines are the level sets of the
statistic $S_X$ or $S_Y$ that the exact optimal transformation depends
on.  In every panel the bands of color run parallel to the black dashed lines:
the recovered transformation is a function of the statistic.  The right
column makes this quantitative by plotting the value of $f$ (and $g$)
against the value of the statistic for all grid points (models 1--3) or
for 3000 configurations sampled from $P$ (model 4): the points collapse
onto a single curve.  For models 1, 2 and 4 this curve is a cosine, as
Eq.~\eqref{eq:synth_rho} predicts: since $f$ is standardized,
$\E[f^2]=1$, its amplitude is $\sqrt2$, and its phase $\alpha$, which
the HGR problem leaves free, is the one selected by the initial point,
so that the maximum need not lie at $S_X=0$.  The obtained $f$ and $g$
collapse onto the same curve with the same phase.
For model 3, whose statistic is not
uniform, the curve is not a cosine, and $g(x_4)$ (green) is a different
function again; both coincide with the exact singular vectors
(Appendix~\ref{app:svdcheck}).
 
As a number, we report the fraction of the variance of $f$ that is
explained by the statistic,
\begin{equation}
  R^2=1-\frac{\E\bigl[\Var[f\,|\,\text{bin of }S_X]\bigr]}{\Var[f]},
  \label{eq:R2}
\end{equation}
where the values of $S_X$ are divided into 40 bins; $R^2=1$ if and only
if $f$ is constant inside every bin, i.e., a function of $S_X$ up to the
bin width.  We find $R^2=1.000000$ for models 1 and 2 (both $f$ and
$g$), $0.9986$ for model 3 and $0.9978$ for model 4.  The deviation from
one in the last two cases is the variation of a smooth $f$ across one
bin, not a dependence on other variables; this is confirmed by the
direct test below.
 
\emph{Irrelevant variables.}  In model 3, the variable $x_2$ is
independent of the others, and in model 4, the variables $x_4,\dots,x_8$
do not enter the relation.  The exact optimal transformation does not
depend on them.  To test the recovered $f$, we take 3000 configurations,
replace the value of one irrelevant variable by a fresh random value,
and record the largest change of $f$ divided by the standard deviation
of $f$.  The result is $2\times10^{-6}$ for $x_2$ in model 3 and exactly
zero for each of $x_4,\dots,x_8$ in model 4.  The middle panel of the
third row of Fig.~\ref{fig:synth_small} shows the same thing as a
picture: $f(x_1,x_2)$ at fixed $x_3$ is constant along $x_2$.  The exact
zero in model 4 has a simple origin: a site whose angle
function is $a_n\equiv0$ has a TT core of $P$ that does not depend on
$x_n$, so the local problem of Sec.~\ref{sec:local} is the same for
every value of $x_n$ and its closed-form solution is constant in $x_n$;
any dependence on $x_n$ in the initial point is removed in the first
sweep.  The rightmost panel of Fig.~\ref{fig:synth_big} shows the response
of $f$ to each variable separately, with the other variables held
fixed.  The five irrelevant variables give flat lines.  The three
relevant ones give three different curves, which are not $\sin x_1$,
$x_2^2/2$ and $x_3$ themselves but these functions seen through the
link $h$: with the other variables fixed,
$f=h\bigl(a_i(x_i)+\text{const}\bigr)$ with
$h(S)=\sqrt2\cos(S-\alpha)$, so that the extrema of the $x_1$ curve
are displaced from those of $\sin x_1$, and the $x_2$ curve,
$\cos(x_2^2/2+\text{const})$, contains a quartic term besides the
quadratic one.  This is the general situation of Appendix~\ref{app:suff}:
the optimal transformation is a function of the sufficient statistic,
but which function of it, the link $h$, is decided by the distribution
(here by the kernel), not by the form in which the statistic was
written, and the algorithm has to find both.  The same point will
reappear in Sec.~\ref{sec:afm}, where $h$ changes from linear to
sign-like with temperature.
 
\emph{Dependence on the bipartition.}  Models 1 and 2 have the same
distribution and the same value $\rho_{\rm TT}=0.726149$, but different
transformations: for model 1 the level sets of $f$ are the lines
$x_1+x_2=$ const (anti-diagonals in Fig.~\ref{fig:synth_small}, top
row), for model 2 they are the lines $x_2-x_4=$ const (diagonals,
middle row).  The optimal transformation is a property of the
distribution \emph{and} of the chosen bipartition. The algorithm
returns the relation that connects the two blocks it was given.
 
\section{Bipartite mean-field antiferromagnet}
\label{sec:afm}
 
The synthetic distributions of the previous section were built so that
their optimal transformations are known.  We now apply the algorithm to
a Boltzmann distribution of a spin model, where the optimal
transformation is not put in by hand, but can still be computed exactly
by a reduction, so that the result can be checked.
 
\subsection{Model and exact reference}
\label{sec:afm_model}
 
Consider $N=2k$ Ising spins $s_i=\pm1$ on the complete bipartite graph: the spins are divided into two groups $A$ and $B$ of $k$ spins
each, every spin of $A$ interacts with every spin of $B$ with the same
antiferromagnetic coupling, and spins in the same group do not
interact.  With $M_A=\sum_{i\in A}s_i$ and $M_B=\sum_{j\in B}s_j$, the
energy is given by
\begin{equation}
  H=\frac{J}{N}\,M_AM_B,\qquad J>0,
  \label{eq:afm}
\end{equation}
and $P(s)\propto e^{-H/T}$.  In the limit $N\to\infty$ it has a transition at $T_c=J/2$ below
which $M_A$ and $M_B$ become macroscopic with opposite signs, and the
staggered magnetization $(M_A-M_B)/N$ is the order parameter.  We take
the two groups as the two blocks of the bipartition, $A$ and $B$, and
ask which function of the spins of $A$ is most correlated with which
function of the spins of $B$.
 
The Boltzmann weight $e^{-H/T}$ has an exact TT representation of bond
dimension $k+1$ when the sites are ordered as
$A_1,\dots,A_k,B_1,\dots,B_k$ with the cut after site $k$.  The bond
index inside block $A$ counts the number of up spins seen so far, so
that the bond at the cut carries $M_A$; each core
in block $B$ multiplies by the factor $e^{-(J/NT)M_As_j}$, which
depends on $M_A$ through the bond index and on the local spin $s_j$, and
passes $M_A$ on to the next site.  For $N=32$ the bond dimension is
$r=17$ and the configuration space has $2^{32}\simeq4\times10^9$ points.

Explicitly, with the bond index $m=0,\dots,k$ counting the up spins
seen so far in block $A$, so that $M_A=2m-k$ at the cut, the cores are
\begin{equation}
  \begin{split}
  W^{(i)}[s_i]_{mm'}&=\delta_{m',\,m+(1+s_i)/2}\quad(i\in A),\\
  W^{(j)}[s_j]_{mm'}&=\delta_{mm'}\,e^{-(J/NT)(2m-k)s_j}\quad(j\in B),
  \end{split}
  \label{eq:afm_cores}
\end{equation}
with the boundary vectors $e_0^{\T}$ (no up spins yet) on the left and
the all-ones vector on the right.  Multiplying out, the product over $A$
selects $m=(k+M_A)/2$ on the cut bond, and the product over $B$ then
gives $\prod_{j\in B}e^{-(J/NT)M_As_j}=e^{-(J/NT)M_AM_B}=e^{-H/T}$.

Since $H$ depends on the configuration only through $(M_A,M_B)$,
the reduction of Appendix~\ref{app:suff} applies: the optimal transformations
are functions of the group magnetizations, $f^{*}=h_f(M_A)$ and
$g^{*}=h_g(M_B)$, and $\rhoHGR$ and the two functions are obtained from
the SVD of the $(k+1)\times(k+1)$ reduced distribution
\begin{equation}
  \tilde P(M_A,M_B)\propto\binom{k}{\frac{k+M_A}{2}}\binom{k}{\frac{k+M_B}{2}}
  e^{-JM_AM_B/NT},
  \label{eq:afm_reduced}
\end{equation}
which is the number of configurations with given $(M_A,M_B)$ times the
Boltzmann factor.  This reduced problem is symmetric under the exchange
of the two groups and under the global spin flip; as a result $h_f$ is an
odd function of $M_A$ and $h_g(M_B)=h_f(-M_B)$, so that the optimal pair
is $f^{*}=h(M_A)$, $g^{*}=h(-M_B)$ with one function $h$.  The algorithm
is not told any of this: it works on the $2^{32}$ configurations through
the TT of Eq.~\eqref{eq:Ptt}, and it is free to return any function of
the sixteen spins of each block.
 
\subsection{Results}
\label{sec:afm_results}
 
\begin{figure*}
\includegraphics[width=\textwidth]{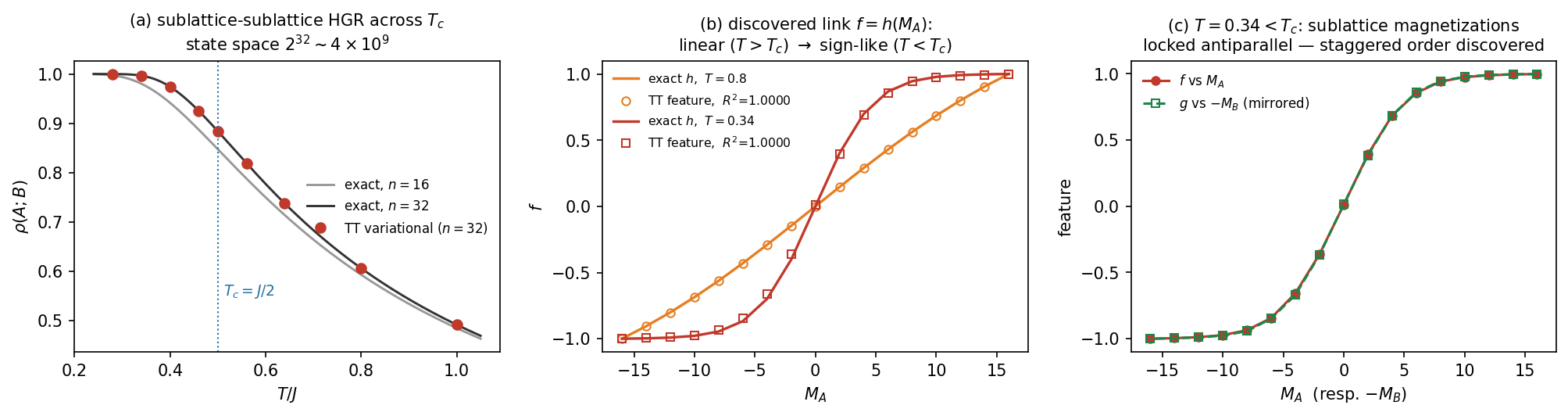}
\caption{\label{fig:afm}
Bipartite mean-field antiferromagnet, $N=32$ spins ($k=16$ per group),
$D=8$.  (a) HGR maximal correlation between the two groups as a function
of temperature: TT algorithm (circles) and exact value from the SVD of
the reduced distribution, Eq.~\eqref{eq:afm_reduced}, for $N=32$ (black)
and $N=16$ (gray).  The dotted line is the mean-field $T_c=J/2$.  (b) Recovered
$f$ against $M_A$ at $T/J=0.80$ and $0.34$ (symbols; mean over 2500
sampled configurations with the given $M_A$) and the exact $h$ (lines),
normalized to $\max|h|=1$; $R^2$ is the fraction of the variance of $f$
explained by $M_A$.  (c) $f$ against $M_A$ and $g$ against $-M_B$ at
$T/J=0.34$.}
\end{figure*}

Figure~\ref{fig:afm} shows the results for $N=32$ ($k=16$) at nine
temperatures between $T/J=0.28$ and $1.0$, with $D=8$, four outer
alternations and a warm start from the linear functions $f_0=M_A$,
$g_0=-M_B$ (rank-2 TTs padded to $D=8$).  The run time is about 5 s per
temperature on a laptop.
 
\emph{Value of the correlation.}  Panel (a) compares $\rho_{\rm TT}$
with the exact value obtained from the SVD of the reduced distribution,
Eq.~\eqref{eq:afm_reduced}.  The two agree to
$4\times10^{-5}$ or better at all temperatures (to $10^{-8}$ at
$T/J=1.0$), and $\rho_{\rm TT}$ is below the exact value at every point.
The correlation between the two groups grows from about $0.5$ at $T=2T_c$
to $1$ as $T\to0$, where the two group magnetizations become locked;
for a finite system the curve is smooth across $T_c$, and the exact
curves for $N=16$ and $N=32$ show the approach to the mean-field limit.
At $N=16$ ($2^{16}$ configurations) the full table of $P$ can also be
built and the SVD of $K$ computed directly; the TT algorithm agrees
with it to $10^{-9}$ in $\rho$ and to six digits in the transformations
at the three temperatures examined (Appendix~\ref{app:svdcheck}).
 
\emph{Form of the transformation.}  To see what function of the spins
the algorithm has found, we sample 2500 configurations from $P$,
evaluate the TT of $f$ on each, and plot the values against $M_A$.
Panel (b) shows the result at $T/J=0.80$ (above $T_c$) and $T/J=0.34$
(below $T_c$) together with the exact $h$ from
Eq.~\eqref{eq:afm_reduced}.  Two facts are visible.  First, at each
temperature all configurations with the same $M_A$ receive the same
value of $f$: the fraction of the variance of $f$ explained by $M_A$,
Eq.~\eqref{eq:R2} with one bin per value of $M_A$, is $R^2=1.0000$ at
both temperatures.  Since $M_A$ takes only $k+1=17$ discrete values, there
is no bin-width effect here, and $R^2=1$ means that $f$ is exactly a
function of $M_A$.  The algorithm has thus found, out of all functions
of sixteen spins, a function of their sum.  Second, the shape of $h$
changes with temperature: above $T_c$ it is close to linear, so that
$f$ is essentially the magnetization $M_A$ itself; below $T_c$ it is
sign-like, saturating at $\pm1$ for $|M_A|\gtrsim 6$, so that $f$ is
essentially the sign of $M_A$, i.e., it tells which of the two
symmetry-related ordered states the group $A$ is in.  The recovered
$h$ coincides with the exact one at both temperatures.
 
Panel (c) shows $f$ against $M_A$ and $g$ against $-M_B$ at $T/J=0.34$.
The two curves fall on top of each other: the function of $B$ that is
most correlated with $f$ is the same function $h$ evaluated at $-M_B$.
The pair $(f,g)$ therefore says that $M_A$ and $M_B$ are locked with
opposite signs, which is the antiferromagnetic (staggered) order, and
it says so without the staggered magnetization having been chosen as an
observable beforehand.
 
\emph{Dependence on $D$ and on the initial point.}  The exact $h(M_A)$
is a function of a sum of $k$ spins and has TT rank $k+1=17$, so the
runs at $D=8$ are truncated.  At $T/J=0.34$ the difference from the
exact value is $1.2\times10^{-4}$ for $D=4$, $3.8\times10^{-5}$ for
$D=8$ (four alternations) and $8.7\times10^{-6}$ with twelve alternations,
$1.4\times10^{-5}$ for $D=12$ and $5.2\times10^{-7}$ for $D=17$ (six
alternations); the bound improves monotonically with $D$ and with
the number of alternations.  Started from random cores instead of the
linear warm start, the algorithm reaches the same curve $h$ but
converges more slowly: at $T/J=0.80$ with $D=8$ the difference from the
exact value is $2\times10^{-4}$ after four alternations and
$1.1\times10^{-4}$ after thirty (three seeds give the same values within
$3\times10^{-5}$).  The warm start used here contains no information
about the answer beyond the first-order statistic $M_A$, which is an
element of the $D=2$ space of Sec.~\ref{sec:setting}; the sign-like shape
of $h$ at low temperature is not in the initial point.
 
\section{Summary and outlook}
\label{sec:summary}

We have shown that the HGR maximal correlation between two subsystems of
a classical many-body system, and the pair of optimal
transformations that carries it, can be computed on the tensor-train
representation of the distribution without ever forming the normalized
matrix $K$ whose SVD defines them.  The transformations are sought as
tensor trains of bond dimension $D$, and the variational problem reduces
to a sequence of closed-form local problems, one per core: the
maximization of a linear functional over an ellipsoid intersected with a
hyperplane, whose metric is the $\PA$-weighted Gram matrix of the frame
matrix and whose inverse performs the division by the marginal core by
core.  Every iterate is a feasible pair, so the objective is a lower
bound on $\rhoHGR$ that increases monotonically, and the cost of a sweep
is linear in the number of variables and polynomial in the bond
dimensions of $P$ and of the transformations.

The algorithm was tested against three kinds of exact reference.  On
systems small enough for the exact SVD of $K$ (Appendix~\ref{app:svdcheck}),
the returned value agrees with $\sigma_2$ to between $10^{-16}$ and
$10^{-8}$ and the transformations with the singular vectors to six
digits, including a case in which the optimal transformation is a
degenerate two-dimensional space.  On synthetic distributions of angles
that hide a nonlinear relation between the two blocks
(Sec.~\ref{sec:synthetic}), the algorithm returns the relation: $f$ and
$g$ are functions of the two statistics that the construction encodes,
nonlinear and non-additive in one case. The results ignore the variables that
do not take part. The obtained $f$ and $g$ depend on the chosen bipartition and not
only on the distribution.  The ten-variable case has $10^{15}$
configurations and no reference other than the closed form.  For the bipartite mean-field
antiferromagnetic model with $2^{32}$ configurations
(Sec.~\ref{sec:afm}), the algorithm returns, out of all functions of
sixteen spins, a function of their sum, $f=h(M_A)$ and $g=h(-M_B)$,
with $h$ linear above $T_c$ and sign-like below it; that is, it returns
the staggered order parameter and, below $T_c$, the sign that tells
which ordered state the block is in, without either having been chosen
beforehand.  In all three cases the comparison also illustrates the
two halves of the statement of Appendix~\ref{app:suff}: the optimal
transformation is a function of the sufficient statistic, but which
function of it, the link $h$, is decided by the distribution, and the
algorithm has to find both.

Several points are left for future work.  (i)~The variational space is
restricted to a single bond dimension $D$, and convergence to a local
maximum of the TT-restricted problem cannot be excluded.  In the
examples studied, the bound improved monotonically with $D$ and with the
number of alternations, and random starts reached the same
transformation as the warm start; a systematic study of the dependence
on $D$, on the least-squares cutoff and on the initial point is not
attempted here.  (ii)~The algorithm computes the leading pair of the
dependence spectrum; several pairs at once could be obtained by a block
version of the sweep with orthogonality constraints, as in the block
TT-SVD of Ref.~\cite{LeeCichocki2015}, or by deflation.  
(iii)~Distributions dominated by hard constraints,
in which most configurations have zero probability, make the local
metric $G$ singular on a large subspace.  The cutoff of
Sec.~\ref{sec:local} removes the null directions, but a systematic
treatment, such as a fugacity regularization followed by the
constrained limit or a restriction of the frame matrix to the support,
is needed before such distributions can be handled routinely.
(iv)~The cost count of Sec.~\ref{sec:local} has not been checked
against measured scaling in $N$, $r$ and $D$.  
(v)~The HGR maximal correlation is one of several measures of the
dependence between two subsystems that are built from the joint
distribution normalized by the product of its marginals; another is the
maximal R\'enyi divergence, which has recently been computed for two
subsystems of a matrix product state by a sweep that solves a generalized
eigenvalue problem \cite{Levin2026}.  There, as here, the product of the
marginals enters as the metric of a variational problem on the tensor
network rather than as an operation on the state.  Whether the two
computations can share their environments, and what the dependence
spectrum says about such divergences, is left open.
(vi)~Finally, the
algorithm takes the distribution as a TT, which may come from a
transfer matrix, from a Hamiltonian through the construction of
Appendix~\ref{app:rotation}, or from a TT fitted to data.  It can therefore be asked of
any such distribution which pair of observables carries the correlation
between two of its subsystems, and where that correlation is weaker or
stronger than a model predicts; the data-driven detection of missing or
extra couplings along these lines is a natural application that we have
not explored.

\begin{acknowledgments}
This work was supported in part by JST FOREST Program (Grant Number JPMJFR216K, Japan).
\end{acknowledgments}

\appendix

\appendix

\section{Proof of Eq.~\eqref{eq:rho=sigma2}}
\label{app:proof}

Let $K=\sum_i\sigma_iu_iv_i^{\T}$ with orthonormal $\{u_i\}$, $\{v_i\}$ and
$\sigma_1=1$, $u_1=\sqrt{\PA}$, $v_1=\sqrt{\PB}$ [Eq.~\eqref{eq:trivial}],
the basis being chosen as in the footnote of Sec.~\ref{sec:svd} in the
degenerate case $\sigma_2=1$.  For unit vectors $u\perp u_1$, $v\perp v_1$ expand
$u=\sum_{i\ge 2}a_iu_i+u_\perp$, $v=\sum_{i\ge2}b_iv_i+v_\perp$, where
$u_\perp$, $v_\perp$ lie in the null spaces of $K^{\T}$ and $K$ respectively.
Then $u^{\T}Kv=\sum_{i\ge 2}\sigma_ia_ib_i\le\sigma_2\sum_{i\ge 2}|a_ib_i|
\le\sigma_2\|a\|\,\|b\|\le\sigma_2$, with equality for $u=u_2$, $v=v_2$.
Translating back with Eq.~\eqref{eq:rescale} proves
Eqs.~\eqref{eq:rho=sigma2} and~\eqref{eq:fstar}.

\section{Convergence of the alternation between $f$ and $g$}
\label{app:power}

This appendix shows that the alternation between $f$ and $g$ of
Sec.~\ref{sec:fixedpoint}, with the updates carried out exactly,
converges to the optimal pair, and at what rate.  The composed relation
\eqref{eq:composed} makes the alternation of the $g$-update
\eqref{eq:gopt} and the $f$-update a power method in the coordinates of
Eq.~\eqref{eq:rescale}.  A $g$-update followed by an $f$-update maps $u$ to
$KK^{\T}u$, divided by its length; iterating a symmetric matrix $M=KK^{\T}$
on a vector in this way is the power method for $M$.  The eigenvectors of
$M$ are the left singular vectors $u_i$ of $K$, with eigenvalues
$\sigma_i^2$.  Expanding the current iterate as $u=\sum_i a_iu_i$, one full
iteration multiplies each coefficient $a_i$ by $\sigma_i^2$ before the
normalization, so that the coefficient of every eigenvector relative to
that of the eigenvector with the largest eigenvalue present in $u$ shrinks
by the factor $(\sigma_i/\sigma_{\rm lead})^2$ per iteration, and the
iterate converges to that leading eigenvector.  
The top eigenvector of $M$ is the trivial mode $u_1$, with eigenvalue
one; but the iteration never acquires a component along it, because
$h_f$ is centered and the normalization in Eq.~\eqref{eq:gopt} keeps it
centered.  In the coordinates of Eq.~\eqref{eq:rescale}, this is the
statement that if $u\perp u_1$ then $K^{\T}u\perp v_1$, since
$v_1^{\T}K^{\T}u=(Kv_1)^{\T}u=u_1^{\T}u=0$. 
The leading eigenvector
available to the iteration is therefore $u_2$, with eigenvalue
$\sigma_2^2$, and the next one is $u_3$.  Hence, starting from any
centered $f$ with a nonzero component along $u_2$, the alternation
converges to $(f^{*},g^{*})$, and the error shrinks by the factor
$(\sigma_3/\sigma_2)^2$ per full iteration, i.e.,  by $\sigma_3/\sigma_2$ per
update, whenever $\sigma_2>\sigma_3$ \cite{Buja1990}.

Two consequences are used in the comparisons with exact references.
First, the rate is set by the gap between $\sigma_2$ and $\sigma_3$,
which is why Table~\ref{tab:svd} of Appendix~\ref{app:svdcheck} reports
$\sigma_2-\sigma_3$.  Second, when $\sigma_2=\sigma_3$ the limit is not
unique but is an element of the degenerate subspace, selected by the
initial point; this is the case for the synthetic distributions of
Sec.~\ref{sec:synthetic}, whose optimal transformation is the
two-dimensional space of Eq.~\eqref{eq:synth_rho}, and the comparison
then uses the projection of Eq.~\eqref{eq:kappa_deg}.  In the TT
algorithm of Sec.~\ref{sec:tt} each update is carried out approximately,
by a single sweep over the cores at fixed bond dimension $D$, so the rate
above applies to the exact alternation; the TT truncation adds to it
the effects examined at the end of Sec.~\ref{sec:afm_results}.

\section{Local problem as a projected least-squares problem}
\label{app:local}

The Lagrange condition of Sec.~\ref{sec:local} only identifies a stationary
point.  That Eq.~\eqref{eq:betasol} is the maximum is seen by a change of
variables.  Let $G=G^{1/2}G^{1/2}$ with $G^{1/2}$ the symmetric square root
on the range of $G$, and set $\gamma=G^{1/2}\beta$, $\tilde c=G^{-1/2}c$,
$\tilde d=G^{-1/2}d$.  In terms of $\gamma$ the constraint
$\beta^{\T}G\beta=1$ becomes $\|\gamma\|=1$, the constraint $d^{\T}\beta=0$
becomes $\tilde d^{\T}\gamma=0$, and the objective becomes
$\tilde c^{\T}\gamma$, so that Problem
\eqref{eq:local_obj}--\eqref{eq:local_const} becomes
$\max_\gamma\tilde c^{\T}\gamma$ over the unit vectors of the hyperplane
$\tilde d^\perp$.  Maximizing a linear functional over the unit vectors of
a hyperplane is elementary: the maximizer is the unit vector along the
projection of $\tilde c$ onto the hyperplane,
\begin{equation}
  \gamma^{*}\propto\tilde c-\frac{\tilde d^{\T}\tilde c}{\tilde d^{\T}\tilde d}\,\tilde d ,
\end{equation}
and the maximum equals the length of that projection.  Multiplying by
$G^{-1/2}$ gives $\beta^{*}\propto G^{-1}c-\lambda G^{-1}d$ with
$\lambda=d^{\T}G^{-1}c/d^{\T}G^{-1}d$, i.e.,  Eq.~\eqref{eq:betasol}, and
the maximum value
\begin{equation}
  J_{\max}=\sqrt{(c-\lambda d)^{\T}G^{-1}(c-\lambda d)} .
  \label{eq:Jmax}
\end{equation}
The two terms $G^{-1}c$ and $G^{-1}d$ in the right-hand side are exactly the two
columns solved for simultaneously in the implementation.

\section{Reduction by sufficient statistics}
\label{app:suff}

This appendix derives a consequence of the self-consistency relation in Eq.~\eqref{eq:fixedpoint} that is used in Secs.~\ref{sec:synthetic}
and~\ref{sec:afm} to obtain exact reference values.  Suppose that the joint
distribution depends on $\vx$ only through a single function
$S(\vx)$ of the configuration of $A$, which may take many fewer values
than $\vx$ itself.  In formulas, there is a function $Q$ of two
arguments such that
\begin{equation}
  P(\vx,\vy)=Q\bigl(S(\vx),\vy\bigr)\qquad\text{for all }\vx,\vy .
  \label{eq:suff}
\end{equation}
For example, if $P$ is the Boltzmann distribution of a spin model whose
energy depends on the spins of $A$ only through their sum, $S(\vx)$ is
that sum.  Then the
marginal $\PA(\vx)=\sum_{\vy}Q(S(\vx),\vy)$ and the conditional
distribution $P(\vy|\vx)=Q(S(\vx),\vy)/\PA(\vx)$ also depend on $\vx$
only through $S(\vx)$: two configurations $\vx$ and $\vx'$ with
$S(\vx)=S(\vx')$ have the same conditional distribution of $\vY$.
Hence the conditional expectation of any function of $\vY$,
\begin{equation}
  \E\bigl[g(\vY)\,\big|\,\vx\bigr]=\sum_{\vy}P(\vy|\vx)\,g(\vy),
  \label{eq:condexp}
\end{equation}
takes the same value at $\vx$ and $\vx'$: for any $g$, the right-hand
side of Eq.~\eqref{eq:condexp} is a function of $S(\vx)$ alone.  Now
substitute $g=g^{*}$.  By Eq.~\eqref{eq:fixedpoint}, the left-hand side of
Eq.~\eqref{eq:condexp} is then $\rhoHGR\,f^{*}(\vx)$, which is therefore
a function of $S(\vx)$ alone as well.  We conclude that the optimal
transformation of $A$ is a function of $S$,
\begin{equation}
  f^{*}(\vx)=h\bigl(S(\vx)\bigr)
  \label{eq:fS}
\end{equation}
for some function $h$ of one variable; the argument does not determine $h$
itself.
In the language of statistics, $S$ is a sufficient
statistic of $\vX$ for $\vY$, and Eq.~\eqref{eq:fS} says that the
optimal transformation never uses more of $\vx$ than the sufficient
statistic; the derivation from Eq.~\eqref{eq:fixedpoint} given here is
self-contained.

If the same holds on the other side, that is, if $P$ depends on $\vy$
only through a function $S'(\vy)$ as well,
$P(\vx,\vy)=\tilde Q(S(\vx),S'(\vy))$, then also $g^{*}(\vy)=h'(S'(\vy))$,
and the whole problem can be rewritten in terms of the pair $(S,S')$.
Let $\tilde P(s,s')=\sum_{\vx:S(\vx)=s}\sum_{\vy:S'(\vy)=s'}P(\vx,\vy)$ be
the joint distribution of the two statistics, obtained by summing $P$
over all configurations with the given values of $S$ and $S'$.  For
$f=h(S)$ and $g=h'(S')$ the mean, the variance and the objective of
Eq.~\eqref{eq:hgr} depend only on $h$, $h'$ and $\tilde P$:
$\E[f]=\sum_s\tilde P_A(s)h(s)$, $\E[f^2]=\sum_s\tilde P_A(s)h(s)^2$ and
$\E[fg]=\sum_{s,s'}\tilde P(s,s')h(s)h'(s')$, where $\tilde P_A$ is the
marginal of $\tilde P$.  Therefore the HGR problem of $(\vX,\vY)$, whose
optimal transformations are of the form $h(S)$ and $h'(S')$, has the
same maximal value as the HGR problem of the pair $(S,S')$ with joint
distribution $\tilde P$, and $h$ and $h'$ are the optimal transformations
of that smaller problem.  When $S$ and $S'$ take few values, $\tilde P$
is a small matrix and $\rhoHGR$, $h$ and $h'$ follow from its SVD
(Sec.~\ref{sec:svd}) at no cost, even when the configuration space of
$(\vx,\vy)$ is enormous.  This is how the reference values of
Secs.~\ref{sec:synthetic} and~\ref{sec:afm} are obtained: in
Sec.~\ref{sec:synthetic}, $S$ and $S'$ are sums of angle functions and
$\tilde P$ depends on them only through their difference, so that its
SVD is a Fourier series; in Sec.~\ref{sec:afm}, they are the
two group magnetizations and $\tilde P$ is a $(k+1)\times(k+1)$ matrix.

\section{Tensor-train form of a function of a sum of single-site terms}
\label{app:rotation}

This appendix gives the TT cores of the distributions of
Sec.~\ref{sec:synthetic} and, more generally, of any weight that depends
on the configuration through a sum of single-site terms,
\begin{equation}
  W(x_1,\dots,x_N)=\Phi\Bigl(\sum_{n=1}^{N}\theta_n(x_n)\Bigr),
  \label{eq:appC_W}
\end{equation}
with $\Phi$ a $2\pi$-periodic function; in Sec.~\ref{sec:synthetic},
$\Phi=C$ and $\theta_n(x_n)=s_na_n(x_n)$.  Expanding $\Phi$ in a Fourier
series, $\Phi(u)=\sum_m\phi_me^{imu}$, and using
$e^{im\sum_n\theta_n}=\prod_ne^{im\theta_n}$,
\begin{equation}
  W(x)=\sum_m\phi_m\prod_{n=1}^{N}e^{im\theta_n(x_n)} ,
  \label{eq:appC_sum}
\end{equation}
a sum of product functions, one per Fourier mode.  A sum of $M$ product
functions is a TT of bond dimension $M$: the bond index labels the mode,
each core is diagonal in it, $W^{(n)}[x_n]_{mm'}=\delta_{mm'}e^{im\theta_n(x_n)}$,
and the coefficients $\phi_m$ are absorbed into one boundary vector.
Since the physical index enters only through the scalar
$e^{im\theta_n(x_n)}$, the bond dimension does not depend on $N$, on the
local dimensions $d_n$, or on the form of the $\theta_n$, which may be
arbitrary nonlinear functions of single variables.

For a real and even $\Phi$, as is the kernel $C$ of Eq.~\eqref{eq:kernel},
$\phi_{-m}=\phi_m$ is real and $\Phi(u)=\phi_0+2\sum_{m\ge1}\phi_m\cos mu$,
and the cores can be made real by replacing the pair of modes
$e^{\pm imS}$ by $(\cos mS,\sin mS)$.  Let $S_n=\sum_{j\le n}\theta_j(x_j)$
be the partial sum up to site $n$.  Since $S_n=S_{n-1}+\theta_n(x_n)$,
the addition formulas for cosine and sine read
\begin{equation}
  \begin{split}
  \bigl(\cos mS_n,\,\sin mS_n\bigr)
  &=\bigl(\cos mS_{n-1},\,\sin mS_{n-1}\bigr)\,R(m\theta_n),\\
  R(\theta)&=\begin{pmatrix}\cos\theta&\sin\theta\\-\sin\theta&\cos\theta\end{pmatrix},
  \end{split}
  \label{eq:appC_rot}
\end{equation}
where $\theta_n$ stands for $\theta_n(x_n)$ here and below; that is,
advancing the partial sum by one site rotates the pair by the angle
$m\theta_n(x_n)$.  Keeping the constant mode and the pairs
$m=1,\dots,J$ on the bond, in the order
$(1,\cos S,\sin S,\cos2S,\sin2S,\dots,\cos JS,\sin JS)$, the cores are
\begin{align}
  W^{(1)}[x_1]&=\bigl(1,\ \cos\theta_1,\ \sin\theta_1,\ \dots,\
    \cos J\theta_1,\ \sin J\theta_1\bigr),
  \label{eq:appC_first}\\
  W^{(n)}[x_n]&=1\oplus R(\theta_n)\oplus R(2\theta_n)\oplus\dots\oplus
    R(J\theta_n),
  \nonumber\\
  &\qquad 2\le n\le N-1,
  \label{eq:appC_mid}\\
  W^{(N)}[x_N]&=\bigl(\phi_0,\ 2\phi_1\cos\theta_N,\ -2\phi_1\sin\theta_N,\
    \dots,
  \nonumber\\
  &\qquad\quad 2\phi_J\cos J\theta_N,\ -2\phi_J\sin J\theta_N\bigr)^{\T},
  \label{eq:appC_last}
\end{align}
where $W^{(1)}$ is a row vector,
$W^{(N)}$ a column vector, and the middle cores are block-diagonal
$(2J+1)\times(2J+1)$ matrices.  Multiplying out from the left, the bond
after site $n$ carries $(1,\cos mS_n,\sin mS_n)_{m=1}^{J}$ by
Eq.~\eqref{eq:appC_rot}, and the last core contracts this vector to
$\phi_0+2\sum_m\phi_m(\cos mS_{N-1}\cos m\theta_N-\sin mS_{N-1}\sin m\theta_N)
=\phi_0+2\sum_m\phi_m\cos mS_N=\Phi(S_N)$, which is Eq.~\eqref{eq:appC_W}.
The bond dimension is $2J+1$ at every bond.  A variable that does not
take part, $\theta_n\equiv0$, has the identity matrix as core, which is
why the cores of $P$ in model~4 of Sec.~\ref{sec:synthetic} do not
depend on $x_4,\dots,x_8$.  A uniform prior $1/d_n$ on each variable is
multiplied into the corresponding core, and the normalization
$Z=\sum_xW(x)$ is obtained by one full contraction and divided into any
one core (Sec.~\ref{sec:setting}); when some variable enters as
$\theta_n(x)=\pm x$ with the uniform prior on the grid, every mode
$m\ne0$ averages to zero over that variable, as in
Sec.~\ref{sec:synth_constr}, and $Z=\phi_0$ times the product of the
prior normalizations.

For the wrapped Gaussian, $\phi_m=c_m=e^{-\sigma^2m^2/2}$, and the
series is truncated at the order $J$ of Eq.~\eqref{eq:trunc}.  The TT
of Eqs.~\eqref{eq:appC_first}--\eqref{eq:appC_last} represents the
truncated kernel exactly, so the algorithm of Sec.~\ref{sec:tt} is run
on an exact TT and the truncation enters only through the definition of
the distribution; with $\sigma=0.8$ and $J=9$ the difference between
the truncated and the full kernel is below $2.5\times10^{-14}$, while
the smallest value of the kernel is $C(\pi)=2.8\times10^{-3}$, so the
distribution stays strictly positive.

The same construction applies to a Boltzmann weight whose energy is a
function of a sum of single-site terms, $e^{-\beta H}$ with
$H=\Phi(\sum_n\theta_n)$.  For a finite set of discrete variables the
sum takes values in a bounded interval, so $e^{-\Phi/T}$ can be
periodized on an interval containing that range and expanded as in
Eq.~\eqref{eq:appC_sum}; the number of modes required is set by the
smoothness of $e^{-\beta\Phi}$ on the interval, and the cores are again
of the form of Eqs.~\eqref{eq:appC_first}--\eqref{eq:appC_last}.  This
is an alternative to constructions that carry the value of the sum
itself on the bond, as in Sec.~\ref{sec:afm_model}, and it is the
natural route when the sum is continuous or takes many values.

\section{Comparison with the exact singular value decomposition on small systems}
\label{app:svdcheck}

For systems small enough that the full joint table can be constructed
($d^N\lesssim 10^{6}$), the results of Sec.~\ref{sec:hgr} provide an exact
reference: $\rhoHGR=\sigma_2$ and $f^{*}=u_2/\sqrt{\PA}$,
$g^{*}=v_2/\sqrt{\PB}$ from the SVD of the normalized distribution $K$ of
Eq.~\eqref{eq:K}.  This appendix compares the TT algorithm with this
reference for the synthetic distributions of Sec.~\ref{sec:synthetic}
(models 1--3) and for the antiferromagnet of Sec.~\ref{sec:afm} at
$N=16$.  We
(i) construct $P$ in the TT format,
(ii) expand it to the full table and compute the SVD of $K$,
(iii) run the algorithm of Sec.~\ref{sec:tt} at bond dimension $D$ and
compare $\rho_{\rm TT}$ with $\sigma_2$, and
(iv) compare the transformations through the $\PA$-weighted correlation
\begin{equation}
  \kappa_f=\Bigl|\,\bigl\langle f_{\rm TT},f^{*}\bigr\rangle_{\PA}\Bigr|
  =\Bigl|\sum_{\vx}\PA(\vx)\,f_{\rm TT}(\vx)\,f^{*}(\vx)\Bigr| ,
  \label{eq:kappa}
\end{equation}
which equals one if and only if $f_{\rm TT}=\pm f^{*}$ (both being
standardized), and similarly $\kappa_g$.  The absolute value removes the
irrelevant common sign of $(f,g)$.  We also report the gap
$\sigma_2-\sigma_3$, since the convergence rate of the alternation between
$f$ and $g$ (Appendix~\ref{app:power}) and the
uniqueness of the optimal transformations depend on it, and a case with
$\sigma_2\approx\sigma_3$ can legitimately return any transformation in the
near-degenerate subspace.

When $\sigma_2$ is exactly degenerate, $\sigma_2=\sigma_3=\dots=\sigma_{m+1}$,
the optimal transformation is not unique: any standardized function in
the $m$-dimensional space spanned by $u_2/\sqrt{\PA},\dots,u_{m+1}/\sqrt{\PA}$
reaches $\rhoHGR$, and the algorithm returns one of them, chosen by the
initial point.  In this case Eq.~\eqref{eq:kappa} is replaced by the
length of the projection of $f_{\rm TT}$ onto this space,
\begin{equation}
  \kappa_f=\Bigl(\sum_{j=2}^{m+1}\bigl\langle f_{\rm TT},
  u_j/\sqrt{\PA}\bigr\rangle_{\PA}^{2}\Bigr)^{1/2},
  \label{eq:kappa_deg}
\end{equation}
which again equals one if and only if $f_{\rm TT}$ is an optimal
transformation.  For $m=1$, Eq.~\eqref{eq:kappa_deg} reduces to
Eq.~\eqref{eq:kappa}.
 
\begin{table*}
\caption{\label{tab:svd}
Comparison of the TT algorithm with the exact SVD of the normalized
distribution $K$ on small systems.  The first three rows are the
synthetic distributions of Sec.~\ref{sec:synthetic}, models 1--3 of
Table~\ref{tab:synth} (the two ``linear'' rows are the same distribution
with two different bipartitions); the last three rows are the bipartite
antiferromagnet of Sec.~\ref{sec:afm} with $N=16$ spins at three
temperatures.  $N$ is the number of sites (for the antiferromagnet, the
number of spins), $k$ the position of the cut, $d$ the local dimension,
$r$ the bond dimension of $P$, and $D$ the bond dimension of $f$ and
$g$.  $\sigma_2-\rho_{\rm TT}$ is the difference between the exact value
and the value returned by the algorithm; it is positive in every row, as
required for a lower bound.  $\kappa_f$ and $\kappa_g$ are the overlaps
of Eqs.~\eqref{eq:kappa} and~\eqref{eq:kappa_deg}; in the two linear
rows $\sigma_2=\sigma_3$ and Eq.~\eqref{eq:kappa_deg} with $m=2$ is
used.}
\begin{ruledtabular}
\begin{tabular}{lcccccccccc}
model & $N$ & $k$ & $d$ & $r$ & $D$ & $\sigma_2$ & $\sigma_2-\rho_{\rm TT}$ &
$\kappa_f$ & $\kappa_g$ & $\sigma_2-\sigma_3$ \\
\hline
model 1, linear, split A & 4  & 2 & 32 & 19 & 6 & 0.726149 & $9\times10^{-16}$ & 1.000000 & 1.000000 & 0 \\
model 2, linear, split B & 4  & 2 & 32 & 19 & 6 & 0.726149 & $1\times10^{-15}$ & 1.000000 & 1.000000 & 0 \\
model 3, cross term      & 4  & 3 & 32 & 25 & 8 & 0.551890 & $4\times10^{-8}$  & 1.000000 & 1.000000 & 0.232 \\
AFM $N=16$, $T/J=0.34$ & 16 & 8 & 2  & 9  & 8 & 0.980173 & $2\times10^{-9}$  & 1.000000 & 1.000000 & 0.344 \\
AFM $N=16$, $T/J=0.50$ & 16 & 8 & 2  & 9  & 8 & 0.847159 & $1\times10^{-9}$  & 1.000000 & 1.000000 & 0.311 \\
AFM $N=16$, $T/J=0.80$ & 16 & 8 & 2  & 9  & 8 & 0.592431 & $2\times10^{-10}$ & 1.000000 & 1.000000 & 0.306 \\
\end{tabular}
\end{ruledtabular}
\end{table*}
 
Table~\ref{tab:svd} summarizes the comparison.  In every case, the value
returned by the algorithm agrees with $\sigma_2$ to between $10^{-16}$
and $10^{-8}$, and the difference $\sigma_2-\rho_{\rm TT}$ is positive
in every row, as it must be for a lower bound.  The size of the
difference has two origins.  In the two linear rows the exact
transformation $\sqrt2\cos(S_X-\alpha)$ is a TT of bond dimension 2
[Eq.~\eqref{eq:synth_rho}], contained in the variational space at $D=6$,
and the difference is at the level of machine precision.  In the other
rows $D$ is smaller than the TT rank of the exact transformation, and
the difference is the truncation error at the given $D$ and number of
alternations; its dependence on both is examined at the end of
Sec.~\ref{sec:afm_results}.
The transformations agree with the exact singular
vectors to six digits, $\kappa_f=\kappa_g=1.000000$, including the two
linear rows, where the optimal transformation is a two-dimensional space
and the algorithm returns one element of it.  The bond dimension $D$ of
$f$ and $g$ in the table is small compared with the bond dimension $r$ of
$P$ and with the exact rank of the optimal transformation (for the bipartite antiferromagnet
rows the exact $f^{*}$ has TT rank $k+1=9$, and $D=8$ is already enough
for a difference of $10^{-9}$).  The distributions in the table were
chosen so that the full table of $P$ ($d^N\le 2^{16}$ and $
32^4\simeq10^{6}$ entries) can still be built; the same distributions
are studied in Secs.~\ref{sec:synthetic} and~\ref{sec:afm} at sizes
where this is impossible.

\bibliography{refs}

\end{document}